\documentclass[lettersize,journal]{IEEEtran}
\usepackage{ dsfont }
\usepackage{cite}
\usepackage{graphicx}
\usepackage{psfrag}
\usepackage{url}
\usepackage{stfloats}
\usepackage{amsfonts,amssymb,amsmath,bm}
\usepackage{array}
\usepackage{caption}
\usepackage{cases}
\usepackage{calc}

\usepackage{longtable}
\usepackage{stfloats}  % Written by Sigitas Tolusis
\usepackage{graphics,booktabs,color,epsfig,enumerate}
\usepackage{multirow}
\usepackage{algorithm,algpseudocode}
\usepackage{subfigure}
\usepackage{epstopdf}
\usepackage{array}
\usepackage{stfloats}
\usepackage{soul}
\usepackage{makecell}
\newtheorem{lemma}{Lemma}
\newtheorem{assum}{Assumption}
\newtheorem{prop}{Proposition}
\newtheorem{Remark}{Remark}
\newtheorem{definition}{Definition}
\begin{document}
	
	\title{Domain Generalization for Cross-Receiver Radio Frequency Fingerprint Identification}
	
	\author{Ying Zhang, Qiang Li$^\dagger$, Hongli Liu, Liu Yang and Jian Yang
		\thanks{This work was supported in part by the National Natural
			Science Foundation of China under Grant 62171110. 
			
			Y. Zhang, Q. Li, H. Liu and L. Yang are with School of Information and Communication Engineering, University of Electronic Science and Technology of China,  Chengdu, P. R. China, 611731. J. Yang is with Beijing Institute of Technology, Beijing, P. R. China, 100081, and also with  Laboratory of Electromagnetic Space
			Cognition and Intelligent Control, Beijing, P. R. China, 100083. Part of this work has been published in IEEE SPAWC 2023~\cite{r15_liu2023receiver}.}
		% <-this % stops a space
		\thanks{$^\dagger$ Corresponding author. E-mail: lq@uestc.edu.cn}% <-this % stops a space
	}
	
	
	
	\maketitle
	
	\begin{abstract}
		Radio Frequency Fingerprint Identification (RFFI) technology uniquely identifies emitters by analyzing unique distortions in the transmitted signal caused by non-ideal hardware. Recently, RFFI based on deep learning methods has gained popularity and is seen as a promising way to address the device authentication problem for  Internet of Things (IoT) systems. However, in \emph{cross-receiver} scenarios, where the RFFI model is trained over RF signals from some receivers but  deployed at a new receiver, the alteration of receiver's characteristics would  lead to data distribution shift and cause  significant performance degradation at the new receiver. To address this problem, we  first perform a theoretical analysis of the cross-receiver generalization error bound  and propose a sufficient condition, named Separable Condition (SC), to minimize the classification error probability on the new receiver.
		Guided by the SC, a Receiver-Independent Emitter Identification (RIEI) model is devised  to   decouple  the  received signals into emitter-related features and receiver-related features and only the emitter-related features are used for identification.  Furthermore,  by leveraging federated learning, we also develop a FedRIEI model to  eliminate the need for centralized collection of raw data from multiple receivers. Experiments on two real-world datasets demonstrate the superiority of our proposed methods over some baseline methods.

	\end{abstract}
	
	
	\section{Introduction}
	\IEEEPARstart{W}{ith} the rapid development of wireless communication technology, the utilization of Internet of Things (IoT) devices and technologies has increased significantly in critical applications such as smart healthcare, smart cities, and intelligent vehicles. However, due to the open nature of wireless channels, their security is more susceptible to illegal eavesdropping and malicious attacks compared to traditional wired networks \cite{r1_sicari2015security}. Therefore, in meeting the growing demand for wireless communication, the pursuit of more efficient and simpler wireless communication security authentication and protection technologies becomes increasingly crucial. Most of the traditional wireless communication networks use cryptographic mechanisms and security protocols to ensure network security, which may not be feasible for wireless IoT devices due to their limited energy and computational resources \cite{r2_yassein2017comprehensive,r3_li2015internet}.
	
	Radio Frequency Fingerprint Identification (RFFI) is an effective identification and authentication technique for wireless communication devices at the physical layer, widely employed in various IoT applications \cite{r4_peng2018design,r5_polak2015identification,r6_zou2016survey}. It distinguishes different devices based on unique radio frequency (RF) fingerprint features within the signals of wireless transmitting devices \cite{r7_soltanieh2020review}. These fingerprint features primarily arise from the non-ideal characteristics of the devices, e.g. D/A, filter, mixer and amplifier, which introduce subtle unintentional modulation in the transmit signal. This unintentional modulation gives rise to unique fingerprint for each emitter, which is difficult to be imitated and tampered by illegal users. Therefore, the RFFI technique can serve as a means of identifying legitimate users and securing wireless communication networks at the physical layer.
	
	The earlier RFFI techniques primarily depend on statistical analysis of features extracted from RF signals, such as signal strength, bandwidth, and phase noise \cite{r8_ellis2001characteristics,r9_hall2003detection,r10_dolatshahi2010identification}. They mainly focused on using manually crafted features for classification, relying heavily on domain knowledge and experts for feature extraction \cite{r11_kennedy2008radio}. Furthermore, the significance of these features is non-deterministic and task-dependent. In recent years,   deep learning (DL) approaches have emerged as a promising tool for RFFI. Compared to traditional methods, DL approaches can capture fine-grained features in RF signals, leading to more accurate and robust identification \cite{r12_shen2022towards,r13_peng2019deep,al2020exposing,r14_shen2023towards}. Deep learning models do not require manual feature engineering, allowing them to learn complex representations of RF signals automatically, making them well-suited for handling complex wireless signals.
	
	In practice, a model may be trained and deployed with data collected from different receivers. However, the impact of the receiver is often overlooked in current RFFI studies. Similar to the emission stage, the non-ideal characteristics of receiver circuits, such as low-noise amplifiers (LNA), analog-to-digital conversions (ADC), matched filters, etc., inevitably introduce signal distortion. This phenomenon is known as receiver fingerprint contamination. Consequently, the data distributions   exhibit significant variations across different receivers.  On the other hand, existing DL-based RFFI models usually require identical distributions for both training and testing stages. As such, the alteration of receiver results in  decline in recognition performance, especially when the trained model needs to generalize to a  new receiver \cite{r15_liu2023receiver,r16_hanna2022wisig}.
	
	To mitigate the influence of receivers, recent studies \cite{r17_zha2023cross,r18_yang2024mitigating,r19_bao2023receiver} have employed both labeled training data and unlabeled test data for joint model training. These methods are categorized as Domain Adaptation (DA), which entails transferring model knowledge from the source domain (i.e. the training receivers) to a target domain (i.e. the new receiver) with  accessible target domain data. However, practical challenges often restrict access to target domain data due to privacy issues, regulatory constraints, or the high costs of data acquisition. Consequently, a more viable approach involves training the model using data from multiple receivers, aiming to develop a cross-receiver RFFI model that can be deployed universally once training is complete. This problem, known as cross-receiver RFFI \cite{r20_shen2023towards,r21_merchant2019toward}, has received scant attention in the current literature. 
	
	Cross-receiver RFFI shares similarities with domain generalization (DG)~\cite{r22_wang2022generalizing}. Inspired by DG, we have developed a cross-receiver RFFI approach, known as the Receiver-Independent Emitter Identification (RIEI) network. The key idea of RIEI is to disentangle the extracted features into the receiver-dependent feature and the emitter-dependent feature via introducing the cross entropy loss, the mutual independence loss and the information entropy loss. These features are then processed such that only the emitter-related features are passed to the subsequent classification network. As a result, we attain a robust RFFI model that is insensitive to the receiver characteristics and has good generalization ability to new receivers.

	The  RIEI model is trained using  centralized data collection and processing, where all RF signals are collected, stored, and processed on a central server. However, practical scenarios involve data originating from various platforms and terminals, which are often stored in a distributed manner. Moreover, uploading data to a central server for centralized learning poses risks of privacy breaches and challenges associated with transmitting large volumes of raw RF data.  To address these challenges,  we also develop a distributed  RIEI model by leveraging federated learning~\cite{r23_mcmahan2017communication}. In experiments, we evaluated RIEI against baseline methods that do not account for cross-receiver effects in a centralized setting. Here, when the inputs are time-domain signals, our proposed method improves accuracy by $10.84\%$ and $9.37\%$ on the HackRF and Wisig real-world datasets, respectively (resp.). When the inputs are time-frequency diagrams, our proposed method improves accuracy by $20.74\%$ and $9.49\%$ on the HackRF and Wisig datasets, resp. In distributed scenarios, compared to baseline methods that disregard receiver impacts, our approach showed $11.24\%$ increase in accuracy, even surpassing the centralized training.
	
	Our contributions are summarized as follows:
	\begin{itemize}
		\item We consider cross-receiver  RFFI, and aim at  developing a  receiver-independent RFFI model, which can effectively address the performance degradation caused by the receivers. And we provide a theoretical analysis of the cross-receiver generalization error bound  and propose a sufficient condition, named Separable Condition (SC), to minimize the classification error probability on the new receiver. 
		\item Guided by SC, we devise a novel RIEI model for achieving cross-receiver RFFI. The RIEI exploits  training  data  from multiple receivers to learn a receiver-independent feature space, thereby eliminating the effect of receivers.
		\item We also developed a distributed  RIEI model, which obviates the need for central collection of raw data from multiple devices, thereby addressing challenges related to distributed data storage, data transmission, and data privacy.
	\end{itemize}
	
	The remainder of the paper is  organized as follows. Sec.~\ref{section:problem_formulation} introduces the cross-receiver RFFI problem. Sec.~\ref{section:RIEI} first conducts a theoretical analysis of cross-receiver RFFI, and then describes the proposed RIEI scheme and its training process. Sec.~\ref{section:FedRIEI} extends RIEI to distributed scenarios. Sec.~\ref{section:Experiments} presents the experimental results. Sec.~\ref{section:Conclusions} concludes the paper.
	
	\section{Problem Formulation}\label{section:problem_formulation}
	
	Let us start with the received signal model of RFFI. During the $\ell$-th time interval, the received signal can be expressed as % \vspace{-5pt}
	\begin{equation} \label{eq:system_model}
		x_\ell(t) = \psi \left(c(t) * \varphi \left(s(t)\cos(\omega_0 t+\theta)\right)\right)+ n(t),
	\end{equation}
	where $(\ell-1)T \leq t\leq \ell T$, $n(t)$ is the noise, $T$ is the length of the interval, $\omega_0$ is the angular frequency, $\theta$ is the initial phase angle, $s(t)$ is random modulating signal, $c(t)$ denotes the channel response,  the operator $*$ denotes convolution, $\varphi(\cdot)$ is a random function modeling nonlinear distortion induced by the emitter's hardware, which encodes the ``fingerprint'' of  the emitter, and  $\psi(\cdot)$ is  a random function modeling the nonlinear reception characteristics induced by the receiver's hardware. Notice that different emitters and receivers have distinct  $\varphi$ and $\psi$, resp.
	
	\begin{figure}[]
		\centering
		\centerline{\includegraphics[width=9cm]{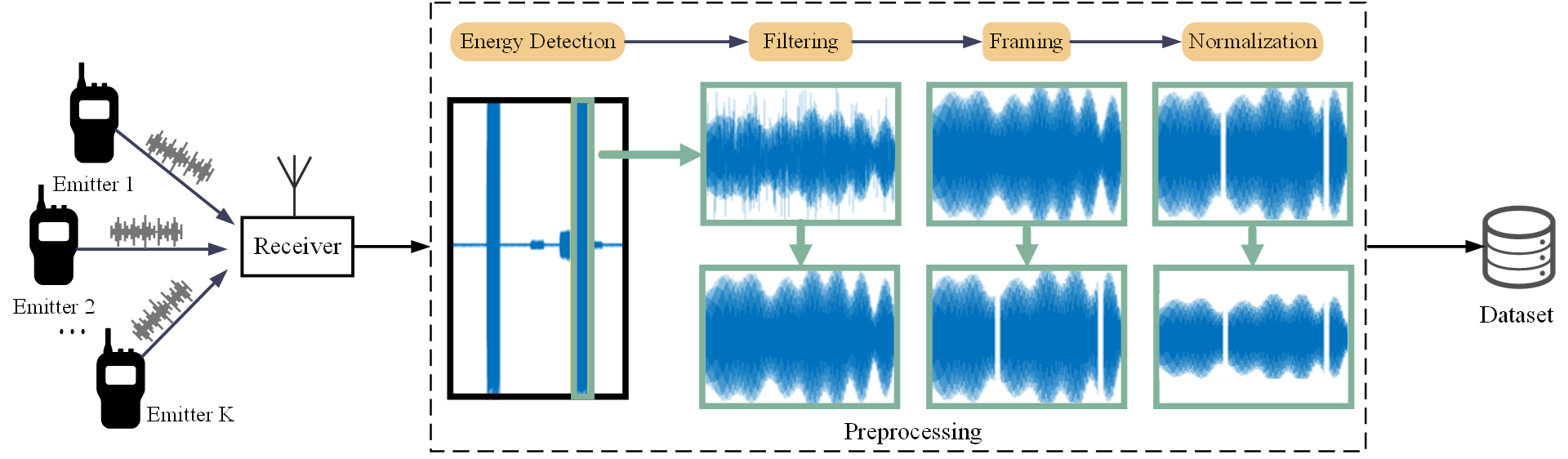}}
		%\vspace{1.5cm}
		\caption{The signal preprocessing flow.}\label{data_preprocessing}
		%\vspace{-5pt}
	\end{figure}
	
	As shown in Fig.~\ref{data_preprocessing}, we usually perform signal preprocessing on the received signal. The preprocessing mainly includes energy detection, filtering, framing and normalization.
	After signal preprocessing, we have a collection of labeled random RF data ${\mathcal S} = \{ {\mathcal S}^1, \ldots, {\mathcal S}^K\}$ from $K$ receivers, and ${\mathcal{S}^k}= \{ (x_{k_i},y_{k_i})\} _{i = 1}^{{N_k}}$ records $N_k$ RF waveform samples $x_{k_i}$ and its emitter index/label $y_{k_i} \in {\mathcal Y}=\{1,\ldots, M\}$, where ${\mathcal Y}$ denotes the set of emitters. Due to receiver fingerprint contamination, the joint distribution $P_{XY}$ varies across the receivers, i.e. $P_{XY}^m \ne P_{XY}^n, 1 \leq m \ne n \leq K$. Our goal is to learn a robust and generalizable classification model $h:\mathcal{X} \to \mathcal{Y}$  that maps the input space to the emitters' index set  to achieve a minimum classification error on   unseen test data $\mathcal{S}^{\mathcal T} = \{ x_{{\mathcal T}_i}\} _{i = 1}^{{N_{\mathcal T}}}$ obtained from a new receiver with $P_{XY}^{\mathcal T} \ne P_{XY}^m$ for $m \in \{ 1, \ldots ,K\} $.
	
\begin{figure}[]
		\centering
		\centerline{\includegraphics[width=9cm]{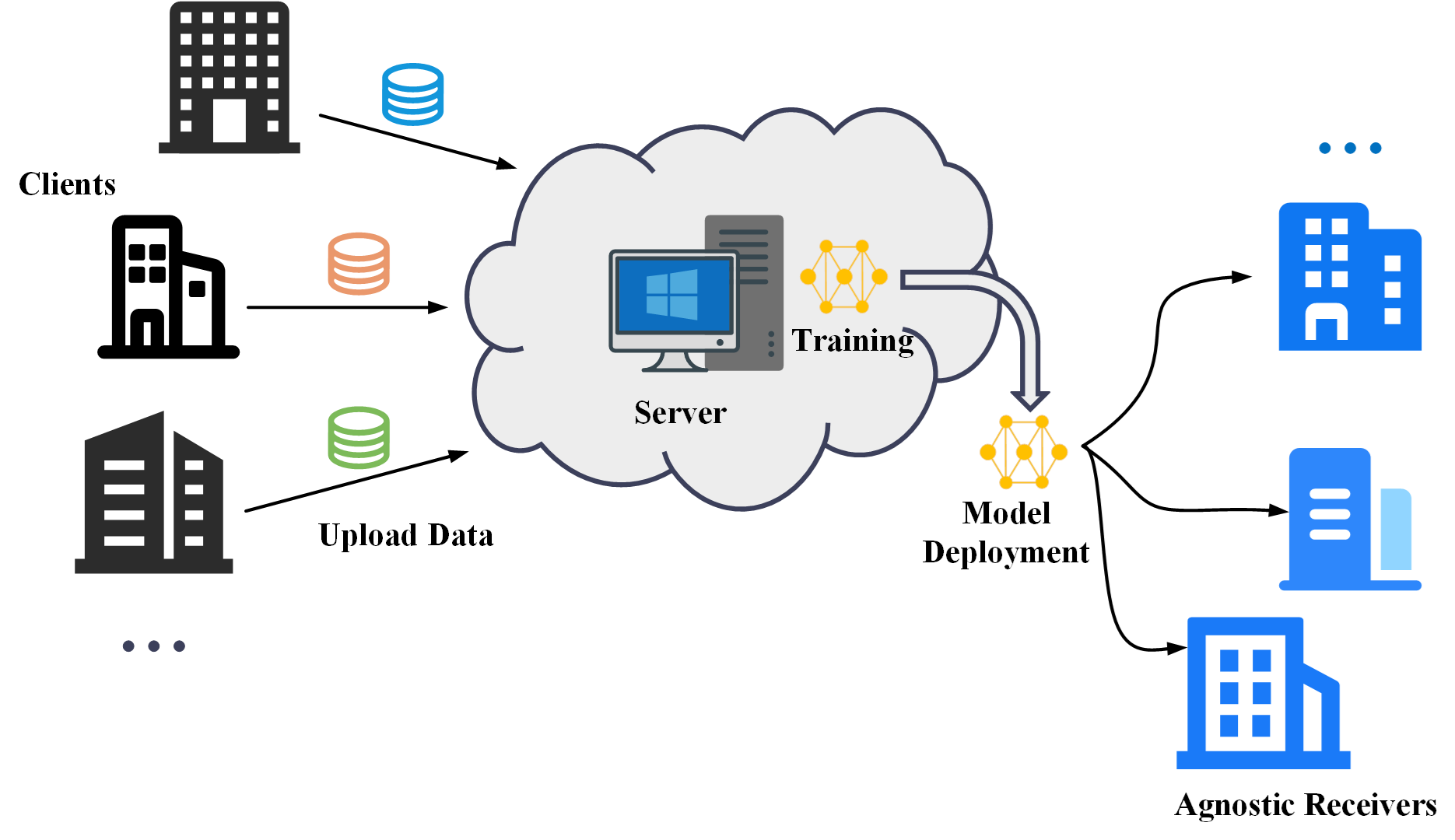}}
		%\vspace{1.5cm}
		\caption{The scenario of cross-receiver RFFI. First, the server collects raw data from multiple receivers. Then, the server utilizes multi-receiver data to train an RFFI model. Finally, the trained model is deployed directly to  new, unseen  receivers.}\label{scenario}
		%\vspace{-5pt}
	\end{figure}

	Now, the multi-source cross-receiver RFFI problem can be described as follows (cf.~Fig.~\ref{scenario}). We are interested in training a classification model $h^\star$ across the clients, which also minimizes the classification error probability on the unseen test domains $\mathcal{S}^{\mathcal T}$, viz.,
	% Mathematically, the receiver-agnostic RFFI problem can be formulated as follows.
	\begin{equation}\label{eq:goal}
		h^\star = \arg\min_{h \in {\cal H}} \epsilon^{\cal T}(h) \triangleq  {\mathbb{E}_{( X,Y) \sim P_{XY}^\mathcal{T}}}[ \mathbb{I} ({h(X) \neq Y})  ]
	\end{equation}
	where $\epsilon^{\cal T}(h)$ denote expected error probability on the test domain, ${\cal H}$ denotes a presumed hypothesis space consisting of the classification models, $\mathbb{E}$ denotes the expectation, $\mathbb{I}(\cdot )$ is the indicator function and  $\mathbb{I}( a\neq b )=1$ if $a\neq b$, and $0$ otherwise.
	% Besides, the proposed solution should follow the same security principle as the traditional FL: only model parameters (e.g., updated gradients) can be sent to the server, and no information about local data can be shared directly. In this paper, each receiver is also named as client.
	
	% Fig.~\ref{1-19}cf.~Fig.~\ref{scenario}
	
	%Now, the multi-source receiver-agnostic RFFI problem can be described as follows (cf.~Fig.~\ref{scenario}). We are interested in training a classification model $h^\star$ across the clients, which also minimizes the classification error probability on the unseen test domains $\mathcal{S}^{\mathcal T}$, viz, 

	Problem~\eqref{eq:goal} is challenging due to the inaccessibility of  the joint distribution $P_{XY}^\mathcal{T}$ or its realization ${\mathcal S}^{\mathcal T}$. In the following, we will provide an alternative way to indirectly minimize the classification error probability over $P_{XY}^\mathcal{T}$.
	
	% Moreover, the federated setting in the receiver-agnostic RFFI scenario poses more difficulties for that. First, the data in FL are stored distributedly, with each client's learning restricted to accessing only its individual local distribution, which constrains to make full use of the multi-source distributions to learn generalizable parameters. Second, though FL has collaborated multi-source data, the local data included in one client naturally form a domain due to the effect of receiver fingerprinting. This leads to different distributions among the collaborative datasets, which is insufficient to ensure domain invariance in a more continuous distribution space to attain good generalizability across other clients. In the following, we will provide an FL-based method to indirectly minimize the classification error probability over $P_{XY}^\mathcal{T}$.

	\section{Proposed Method}\label{section:RIEI}
	\subsection{Theoretical Analysis}
	We first make a theoretical analysis of the cross-receiver RFFI problem, which will also shed light on the subsequent method development.
	
	The following lemma characterizes the generalization performance in the target domain.

	\begin{lemma}
		% {\cite[Theorem 2]{ben2010theory}}
		\label{target_error_bound}
		Let $\mathcal{H}$ be a hypothesis space of VC dimension $d$. If $\mathcal{S}$, $\mathcal{T}$ contains samples of size $N$ each, then for any $\rho \in (0,1)$, with probability at least $1-\rho$ (over the choice of the samples), for every $h \in \mathcal{H}$ the following inequality holds
		\begin{equation}
			\begin{aligned}
				{\epsilon}^t(h) \le ~&{\epsilon}^s(h) + \frac{1}{2}d_{\mathcal{H}\Delta\mathcal{H}}({\cal D}^s, {\cal D}^t) + \Lambda,
			\end{aligned}
		\end{equation}
		%	$f^s: \mathcal{X} \to \mathcal{Y}$ and $f^t: \mathcal{X} \to \mathcal{Y}$ are the ground-truth labeling functions on the source domain and the target domain, respectively,
		where ${\epsilon}^t(h)$ and ${\epsilon}^s(h)$ denote expected error probability on the source and target domain, resp., $\Lambda = 4 \sqrt{\frac{2d\log{2N}+\log{\frac{2}{\rho}}}{N}} + \Lambda^\star$, $\Lambda^\star = \min_{h \in \mathcal{H}}{\epsilon^s(h) + \epsilon^t(h)}$ denotes the minimum combined error probability on both domains, $d_{\mathcal{H} \Delta \mathcal{H}}({\cal D}^s, {\cal D}^t) = 2\sup_{{h}, {h'} \in \mathcal{H}} { | \mathbb{E}_{{X}\sim {\cal D}^s}[|{h}({X}) - {h'}({X})|] - \mathbb{E}_{{X}\sim {\cal D}^t}[|{h}({X}) - {h'}({X})|] }|$ represents the  discrepancy of $\mathcal{S}$ and $\mathcal{T}$ with respect to  $\mathcal{H}$, ${\cal D}^s$ and ${\cal D}^t$ represent the source and target domain, resp.
	\end{lemma} 
	
	From Lemma~\ref{target_error_bound}, we see that  to minimize the classification error probability  on the unseen target domain, we should make the source domain risk  ${\epsilon}^s(h)$ and  the  domain discrepancy $d_{\mathcal{H}\Delta\mathcal{H}}({\cal D}^s, {\cal D}^t)$  as small as possible. For the former,  it can be attained by standard supervised learning with labeled source training data. However, for the latter, if we do not impose any constraint on the target domain, the domain discrepancy could be arbitrarily large. In light of this, we make the following model assumption on data generalization to facilitate the analysis.    
	
	\begin{assum}[Domain transformation model~\cite{r24_robey2021model}]\label{assum:1}
		Let $X$ denote the emitter-related variable and $\nu$ denote receiver/domain-related variable.  We assume that there exists a measurable function $G: {\cal X} \times { \nu} \rightarrow {\cal X}$, referred to as a domain transformation model, which parameterizes the inter-domain covariate shift via
		\begin{equation}
			\mathbb{P}({X^\nu}) =^d G \# \left(\mathbb{P}(X) \times \mathbb{P}(\nu) \right) ,
		\end{equation}
		where $\#$ denotes the push-forward measure and $=^d$ denotes equality in distribution.
	\end{assum}
	
	In addition, we introduce the following definition.
	
	\begin{definition}[Separable Condition]\label{def:sep_cond}
		Let $\tilde{\cal D}$ be feature space associated with $\cal D$ and    $\tilde{h}$  be an equivalent learning model of $h\in {\cal H}$  in the feature space, i.e. $h(X^\nu) = \tilde{h}(Z)$ for some $Z\in \tilde{\cal D}$. Then, $X^\nu$ is called separable under  ${\cal H}$ if for any $X^\nu \in {\cal D}$ and ${h} \in {\cal H}$, the  $Z\in \tilde{\cal D}$ admits a decomposition  $Z = [Z_X, Z_\nu]$, which  satisfies
		\begin{enumerate}
			\item $\nu \rightarrow X \rightarrow Z_X$, i.e. given $X$, $Z_X$ is independent of $\nu$.
			\item $X \rightarrow \nu \rightarrow Z_\nu$, i.e. given $\nu$, $Z_\nu$ is independent of $X$.
			\item $Z_X \perp Z_\nu$, i.e., $Z_X$ is statistically independent of $Z_\nu$.
		\end{enumerate} 
	\end{definition}
	
	The following result characterizes $d_{\mathcal{H}\Delta\mathcal{H}}({\cal D}^s, {\cal D}^t)$ under the transformation model.
	\begin{prop}\label{prop:1}
		Suppose that Assumption~\ref{assum:1} holds and $X^\nu$ satisfies the separable condition on both ${\cal D}^s$ and ${\cal D}^t$. Then, 
		\begin{equation} \label{eq:dis_zero}
			d_{\mathcal{H}\Delta\mathcal{H}}({\cal D}^s, {\cal D}^t) = 0
		\end{equation}
		if $h(X^\nu) = \tilde{h}(Z_X), ~\forall X^\nu \in \{{\cal D}^s, {\cal D}^t \}$.
	\end{prop}
	
	Before presenting the proof of Proposition~\ref{prop:1}, we remark that the condition $h(X^\nu) = \tilde{h}(Z_X)$ can be easily  fulfilled by just discarding the emitter-irrelevant feature $Z_{\nu}$ and making prediction only based on $Z_X$. 
	
	{\it Proof.}~
	\begin{equation} \label{eq:dis_s_t}
		\begin{aligned}
			& d_{\mathcal{H}\Delta\mathcal{H}}({\cal D}^s, {\cal D}^t)  \\
			= &  2\sup_{{h}, {h'} \in \mathcal{H}}  \Big| \mathbb{E}_{{X^{\nu_s}}\sim {\cal D}^s}\big|{h}({X^{\nu_s}}) - {h'}({X^{\nu_s}})\big| \\
			& - \mathbb{E}_{{X^{\nu_t}}\sim {\cal D}^t} \big|{h}({X^{\nu_t}}) - {h'}({X^{\nu_t}}) \big| \Big|  \\
			= & 2\sup_{{h}, {h'} \in \mathcal{H}}  \Big| \mathbb{E}_{{Z}\sim \tilde{\cal D}^s}\big|{h}({X^{\nu_s}}) \\
			& - {h'}({X^{\nu_s}})\big| - \mathbb{E}_{Z \sim \tilde{\cal D}^t} \big|{h}({X^{\nu_t}}) - {h'}({X^{\nu_t}}) \big| \Big|  \\
			= & 2\sup_{{h}, {h'} \in \mathcal{H}}  \Big| \mathbb{E}_{{Z_X, Z_{\nu_s}}\sim \tilde{\cal D}^s}\big|{h}({X^{\nu_s}}) - {h'}({X^{\nu_s}})\big| \\
			& - \mathbb{E}_{Z_X, Z_{\nu_t} \sim \tilde{\cal D}^t} \big|{h}({X^{\nu_t}}) - {h'}({X^{\nu_t}}) \big| \Big|.  
		\end{aligned}
	\end{equation}
	Notice that 
	\begin{equation}\label{eq:dis_source}
		\begin{aligned}
			& \mathbb{E}_{{Z_X, Z_{\nu_s}}\sim \tilde{\cal D}^s}\big|{h}({X^{\nu_s}}) - {h'}({X^{\nu_s}})\big| \\
			= & \mathbb{E}_{X,\nu_s} \mathbb{E}_{{Z_X, Z_{\nu_s}| X,\nu_s}} \big|{h}({X^{\nu_s}}) - {h'}({X^{\nu_s}})\big| \\
			= & \mathbb{E}_{\nu_s} \mathbb{E}_{X} \mathbb{E}_{{Z_X, Z_{\nu_s}| X,\nu_s}} \big|{h}({X^{\nu_s}}) - {h'}({X^{\nu_s}})\big| \\
			= & \mathbb{E}_{\nu_s} \mathbb{E}_{X} \mathbb{E}_{{Z_{\nu_s}|\nu_s}} \mathbb{E}_{{Z_X| X}}  \big|{h}({X^{\nu_s}}) - {h'}({X^{\nu_s}})\big|\\
			= & \mathbb{E}_{{Z_{\nu_s},\nu_s}} \mathbb{E}_{{Z_X, X}}  \big|{h}({X^{\nu_s}}) - {h'}({X^{\nu_s}})\big|,
		\end{aligned} 
	\end{equation}
	where the third equality is due to separable condition. Moreover, if $h(X^\nu) = \tilde{h}(Z_X), ~\forall X^\nu \in \{{\cal D}^s, {\cal D}^t \}$, then \eqref{eq:dis_source} can be further expressed as
	\begin{equation}\label{eq:dis_s}
		\begin{aligned}
			& \mathbb{E}_{{Z_X, Z_{\nu_s}}\sim \tilde{\cal D}^s}\big|{h}({X^{\nu_s}}) - {h'}({X^{\nu_s}})\big| \\
			= &  \mathbb{E}_{{Z_{\nu_s},\nu_s}} \mathbb{E}_{{Z_X, X}}  \big|{\tilde{h}}(Z_X) - {\tilde{h}'}(Z_X)\big| \\
			=  & \mathbb{E}_{{Z_X, X}}  \big|{\tilde{h}}(Z_X) - {\tilde{h}'}(Z_X)\big|. 
		\end{aligned} 
	\end{equation}
	Similarly, we have
	\begin{equation}\label{eq:dis_t}
		\begin{aligned}
			& \mathbb{E}_{{Z_X, Z_{\nu_t}}\sim \tilde{\cal D}^t}\big|{h}({X^{\nu_t}}) - {h'}({X^{\nu_t}})\big| \\
			= & \mathbb{E}_{{Z_X, X}}  \big|{\tilde{h}}(Z_X) - {\tilde{h}'}(Z_X)\big|.
		\end{aligned} 
	\end{equation}
	By substituting \eqref{eq:dis_s} and \eqref{eq:dis_t} into \eqref{eq:dis_s_t}, we arrive at  \eqref{eq:dis_zero}, which completes the proof. \hfill $\blacksquare$
	
	\begin{Remark}
		From   Proposition~\ref{prop:1}, we see that under the premise of the domain transformation model (Assumption~\ref{assum:1}), null  discrepancy can be attained if we can learn a model with feature  satisfying separable condition on ${\cal D}^s$ and ${\cal D}^t$. However, since we cannot access ${\cal D}^t$, the best we can do is to make separable condition valid over ${\cal D}^s$.
	\end{Remark}

	\begin{figure}[]
		\centering
		\centerline{\includegraphics[width=9cm]{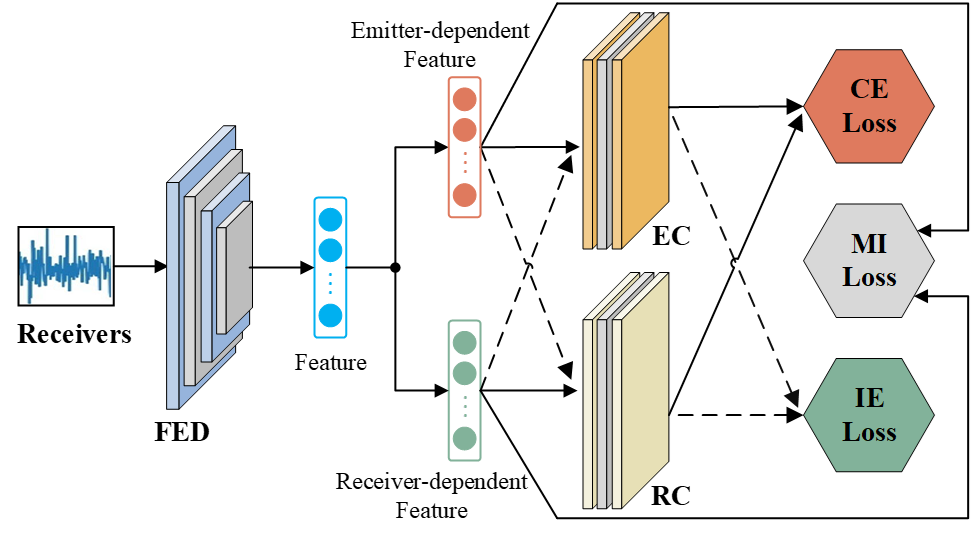}}
		%\vspace{1.5cm}
		\caption{Illustration of the proposed RIEI method.}\label{model}
		%\vspace{-5pt}
	\end{figure}

	\subsection{Overview of The Proposed RIEI Network}
	Guided by Proposition~\ref{prop:1}, the proposed Receiver-Independent Emitter Identification (RIEI) network is shown in Fig.~\ref{model}. It  consists of three main components, namely the Feature Extraction and Disentanglement (FED) module, the Emitter Classifier (EC) and the Receiver Classifier (RC). 
	
	The FED module  maps the input RF waveform into a high-dimensional feature space such that the  emitter-dependent feature (corresponding to $Z_X$) and the receiver-dependent feature (corresponding to $Z_\nu$) are disentangled.
	% To this end, some judiciously designed loss functions and training procedure are needed; we will detail this shortly.

	The EC module  maps the emitter-dependent feature into an $M$-dimensional probability vector, which provides the confidence for predicting which emitter the input signal is from.

	The RC module  has a similar function as the EC module, except that the input is the  receiver-dependent feature and the output  is a $K$-dimensional vector, which provides the confidence for predicting which receiver the input signal is from. Note that after training, the RC module is discarded and only the FED and EC modules are used during the test stage.
	
	These modules together with some judiciously designed training loss functions promote to learn a feature space $Z$  satisfying Separable Condition in Definition~\ref{def:sep_cond}. Specifically,  three types of loss functions are considered:
	\begin{enumerate}
		\item The classification entropy (CE) loss, which exploits the emitters'   and the receivers' label information to minimize  classification error probability over $\cal S$.
		\item The mutual independence (MI) loss, which promotes   independence  between the emitter-dependent feature and the receiver-dependent feature.
		\item The information entropy (IE) loss, which is obtained by calculating the entropy of the EC (RC) module's output when the receiver-dependent (emitter-dependent) feature is input. By maximizing the IE loss, we hope that the EC (RC) module cannot distinguish the receivers (emitters), and thus the emitters' information is not leaked to the receiver-dependent feature, and vice versa.
	\end{enumerate}

	We will explain these losses in detail in the sequel.

	\subsection{The Loss Functions}
	%In the following, we will give more details on the specially designed loss functions. 
	To proceed, let us denote by $\phi_{\theta_F}(\cdot)$, $\phi_{\theta_E}(\cdot)$ and $\phi_{\theta_R}(\cdot)$ the network  function associated with the FED, the EC and RC modules, resp., and  $\theta_F \in \mathbb{R}^{L_F}$, $\theta_E \in \mathbb{R}^{L_E}$ and $\theta_R \in \mathbb{R}^{L_R}$ are the corresponding network parameters. Given an input signal $x$, the FED output vector is denoted as 
	\begin{equation}
		\phi_{\theta_F}(x) \triangleq [\phi_{\theta_F}^{(E)}(x), ~\phi_{\theta_F}^{(R)}(x)].
	\end{equation}
	Our goal is to learn a representation $	\phi_{\theta_F}(x) $ so that the first part of $	\phi_{\theta_F}(x) $, i.e.   $\phi_{\theta_F}^{(E)}(x)$, corresponds to the emitter-dependent features, while the last part of $	\phi_{\theta_F}(x) $, i.e. $\phi_{\theta_F}^{(R)}(x)$, corresponds to the receiver-dependent features. This is achieved through the following loss functions.
	
%	To this end, we will design several loss functions, as explained below.
	
%	where $\phi_{\theta_F}^{(E)}(x)$ and $\phi_{\theta_F}^{(R)}(x)$ represent the emitter-dependent features and the receiver-dependent features {\color{blue}separated  by dimension}, respectively.
	\subsubsection{Classification Entropy Loss}\label{ceo}
	The classification Entropy Loss is motivated by conditions 1) and 2) of the Separable Condition, where the feature $Z_\nu$ (resp.  $Z_X$) should be exclusively dependent on receiver (resp. emitter), containing no information related to the emitter (resp. receiver). This specificity renders $Z_\nu$ (resp.  $Z_X$) as effective statistics for receiver (resp. emitter) classification. 
	
	%The classification here includes two parts, one for classifying emitters and the other for classifying receivers. Notably, 
	
	%To proceed, let us denote by $\phi_{\theta_F}(\cdot)$, $\phi_{\theta_E}(\cdot)$ and $\phi_{\theta_R}(\cdot)$ the network  function associated with the FED, the EC and RC modules, respectively, and  $\theta_F \in \mathbb{R}^{L_F}$, $\theta_E \in \mathbb{R}^{L_E}$ and $\theta_R \in \mathbb{R}^{L_R}$ are the corresponding network parameters. Given an input signal $x$, the FED output is expressed as 
	%\begin{equation}
	%\phi_{\theta_F}(x) = [\phi_{\theta_F}^{(E)}(x), ~\phi_{\theta_F}^{(R)}(x)],
	%\end{equation}
	%where $\phi_{\theta_F}^{(E)}(x)$ and $\phi_{\theta_F}^{(R)}(x)$ represent the emitter-dependent feature and the receiver-dependent feature, respectively.
	
	To ensure accurate classification, by passing the $\phi_{\theta_F}^{(E)}(x)$ to the emitter classifier, we have the output
	\begin{equation}
		z^E = \sigma \left( \phi_{\theta_E} \left(\phi_{\theta_F}^{(E)}(x) \right) \right) \in \mathbb{R}_+^M,
	\end{equation}
	where $ \sigma(\cdot)$ denotes the softmax function, i.e. $[{ \sigma}({x})]_m = \frac{e^{x_m}}{\sum_{j=1}^{M} {e^{x_j}}} $ for $x \in  \mathbb{R}^M$. Notice that $z^E$ lies in a probabilistic simplex satisfying $[z^E]_m \geq 0, \sum_{m=1}^M [z^E]_m =1 $. To encourage the EC correctly classifies the emitters, the cross-entropy loss is  resorted:
	\begin{equation}\label{eq:CE_E}
		\mathcal{L}_{CE}^{E} =   \sum_{k=1}^K \sum_{(x,y) \in {\mathcal S}^k} {\sf H}(z^E, \mathds{1}_{m}),	
	\end{equation}
	where $\sf H(\cdot, \cdot)$ denotes the cross-entropy  function $\mathsf{H}({ p},  { q}) = - \sum_{i} q_i \log{p_{i}}$, and $\mathds{1}_m \in \mathbb{R}^M$ is a one-hot vector with the $m$-th element being one and zeros otherwise.
	
	Similarly, we can calculate the cross-entropy loss for classifying the receivers via the RC module as follows. 
	\begin{equation}\label{eq:CE_R}
		\mathcal{L}_{CE}^{R} =  \sum_{k=1}^K \sum_{(x,y) \in {\cal S}^k} {\sf H}(z^R, \mathds{1}_k),
	\end{equation}
	where $z^R = \sigma \left( \phi_{\theta_R} \left(\phi_{\theta_F}^{(R)}(x) \right) \right) \in \mathbb{R}_+^K,$ and the one-hot vector $\mathds{1}_k \in \mathbb{R}^K$.
	
	By combining~\eqref{eq:CE_E} and \eqref{eq:CE_R}, we get the classification entropy loss:
	\begin{equation} \label{eq:CE}
		\mathcal{L}_{CE} = \mathcal{L}_{CE}^{E} + \mathcal{L}_{CE}^{R}.
	\end{equation}

	\subsubsection{Information Entropy Loss}\label{ie}
	The information entropy loss is also motivated by conditions 1) and 2) of the Separable Condition. Specifically, given an emitter, $Z_X$ is independent of $\nu$, containing no information about the receivers. That is, when feeding $Z_X$ into the RC module, the RC cannot correctly identify which receiver the $Z_X$ is from, and the classification output resembles uniform distribution. From an information theoretic point of view, the classification output attains maximum information entropy. Similarly, when feeding $Z_\nu$ into EC module, the classification output should also attain maximum information entropy. In light of this, we propose the following information entropy (IE) loss ${\mathcal L}_{{{IE}}}$ to encourage feature disentanglement.
	%the emitter-dependent feature should contain little information about the receivers. That is,  when feeding $Z_X$ into the RC module,  the output distribution across various receivers for a specific emitter is close to uniform distribution, which indicates that the predictions made by the RC will possess maximum uncertainty, rendering it unable to accurately predict the receiver based on $Z_X$. That is, the larger the ${\mathcal L}_{IE}^{E\rightsquigarrow R}$, the greater the uncertainty in the prediction made by the RC. Consequently, we propose to {\it maximize} the IE loss ${\mathcal L}_{{{IE}}}$ to further encourage feature disentanglement.
	% ${\cal L}_{{{IE}}}$ is denoted as
	\begin{equation}\label{eq:DE_def}
		{\mathcal L}_{{{IE}}} = {\mathcal L}_{{{IE}}}^{{{R\rightsquigarrow E}}} + {\mathcal L}_{{{IE}}}^{{{E\rightsquigarrow R}}},
	\end{equation}
	where ${\mathcal L}_{{{IE}}}^{{{R\rightsquigarrow E}}}$ calculates the output entropy when the receiver-dependent feature is input into the emitter classifier, i.e., %\vspace{-3pt}
	\begin{equation}\label{eq:classen}
		\mathcal{L}_{{{IE}}}^{{{R\rightsquigarrow E}}} = -\sum_{k=1}^K \sum_{(x,y)\in {\mathcal S}^k }  z^{R\rightsquigarrow E} \odot \log\left( z^{R\rightsquigarrow E} \right),
	\end{equation}
	where $z^{R\rightsquigarrow E} =\sigma \left( \phi_{\theta_E} \left(\phi_{\theta_F}^{(R)}(x) \right) \right) \in \mathbb{R}_+^M$, the $\log(\cdot)$ operation is taken on every element of its argument,  and $\odot$ denotes the element-wise product. ${\mathcal L}_{{{IE}}}^{{{E\rightsquigarrow R}}}$ calculates the output entropy when the emitter-dependent feature is input into the receiver classifier, i.e., 
	\begin{equation}\label{eq:domainen}
		\mathcal{L}_{{{IE}}}^{{{E\rightsquigarrow R}}} = -\sum_{k=1}^K \sum_{(x,y)\in {\mathcal S}^k}z^{E\rightsquigarrow R}  \odot \log\left(  z^{E\rightsquigarrow R} \right),
	\end{equation}
	where $z^{E\rightsquigarrow R} =\sigma \left( \phi_{\theta_R} \left(\phi_{\theta_F}^{(E)}(x) \right) \right)\in \mathbb{R}_+^K$.
	
	%The rationale behind the design of the IE loss ${\mathcal L}_{{{IE}}}$ is as follows. Take ${\mathcal L}_{IE}^{E\rightsquigarrow R}$ for example. According to condition 2) in Definition 1, given an emitter, $Z_X$ is independent of $\nu$. From information theoretic point of view, the emitter-dependent feature should contain little information about the receivers. That is,  when feeding $Z_X$ into the RC module,  the output distribution across various receivers for a specific emitter is close to uniform distribution, which indicates that the predictions made by the RC will possess maximum uncertainty, rendering it unable to accurately predict the receiver based on $Z_X$. That is, the larger the ${\mathcal L}_{IE}^{E\rightsquigarrow R}$, the greater the uncertainty in the prediction made by the RC. Consequently, we propose to {\it maximize} the IE loss ${\mathcal L}_{{{IE}}}$ to further encourage feature disentanglement.
	
	\subsubsection{Mutual Independence Loss}\label{olo}
	The mutual independence loss is motivated by condition 3) of the Separable Condition, i.e., $Z_X\perp Z_\nu$.  Proposition~\ref{prop:1} reveals that  to minimize the domain discrepancy $d_{\mathcal{H} \Delta \mathcal{H}}({\cal D}^s, {\cal D}^t)$, it is plausible to  decouple the feature space  into two  mutually independent  subspaces, one for   emitter-related feature  $Z_X$ and the other for receiver-related feature $Z_\nu$. As such, we propose the following mutual independence loss:
	% to learn a feature space satisfying the 1st condition in the Separable Condition definition.
	%To reduce the expected error probability in the target domain, we then focus on minimizing the domain discrepancy $d_{\mathcal{H} \Delta \mathcal{H}}({\cal D}^s, {\cal D}^t)$. When condition 1) in Definition 1 is satisfied, i.e., the emitter-related feature $Z_X$ and the receiver-related feature $Z_\nu$ are statistically independent, a reduction in the discrepancy $d_{\mathcal{H} \Delta \mathcal{H}}({\cal D}^s, {\cal D}^t)$ between the source domain and target domain  is anticipated. Therefore, with the aid of the high expressiveness of the neural networks, our efforts are directed toward ensuring the statistical independence of $Z_X$ and $Z_\nu$ within specific high-dimensional feature spaces. 
	%
	%To encourage the FED network to learn a receiver-independent feature space, we introduce the following loss to force mutual independence between $\phi_{\theta_F}^{(E)}(x)$ and $\phi_{\theta_F}^{(R)}(x)$. 
	\begin{equation}\label{eq:orth}
		{\mathcal{L}_{{\text{MI}}}} =  \sum_{k=1}^K \sum_{(x,y) \in {\cal S}^k} \frac{\left| {\langle \phi _{{\theta _F}}^{(E)}(x),\phi _{{\theta _F}}^{(R)}(x)\rangle } \right|}{\|\phi_{\theta_F}^{(E)}(x) \| \|\phi_{\theta_F}^{(R)}(x)\|},
	\end{equation}
	where $\| \cdot \|$ denotes the Frobenius norm, and $\langle\cdot,\cdot \rangle$ denotes the inner product of two vectors.

	\subsection{Training}
	The overall loss function is given by	
	% \vspace{-5pt}
	\begin{equation}\label{eq:overall_loss}
		{\mathcal L}(\theta_F,\theta_E, \theta_R) = {\mathcal L}_{CE} + \lambda_1 {\mathcal L}_{MI} - \lambda_2 {\mathcal L}_{IE},
		% \vspace{-5pt}
	\end{equation}
	where ${\mathcal L}_{CE}$, ${\mathcal L}_{IE}$ and ${\mathcal L}_{MI}$ are given in Eq.~\eqref{eq:CE}, \eqref{eq:DE_def} and \eqref{eq:orth}, resp.; $\lambda_i \in \mathbb{R}_+, ~i=1,2$ are hyperparameters. A straightforward way to train $(\theta_F,\theta_E, \theta_R)$ is using stochastic gradient descent (SGD) to update the parameters simultaneously. However, we numerically found that SGD  is not stable and usually results in poor performance for our considered loss. We guess the reason is that the feature disentanglement is much harder than conventional feature extraction task, and the former is more sensitive to the update of the classifier. Therefore, the update of the classifier and the FED should be carefully balanced. To this end, we propose an alternating training with intermediate updating scheme to optimize the network parameters. Specifically, it includes the following steps. Randomly initialize $(\theta_F^0,\theta_E^0, \theta_R^0)$  and alternately update the classifier and the feature extractor:
	\begin{enumerate}
		\item Update the classifier: \vspace{-4pt}
		\begin{subequations} \label{eq:train}
			\begin{align}
				\theta_E^{\ell+1} &  \leftarrow   \theta_E^{\ell}   -  \eta_E^{\ell} \frac{\partial {\cal L}_{CE}}{\partial \theta_E} \label{eq:train_a}, \\
				\theta_R^{\ell+1} &  \leftarrow   \theta_R^{\ell}   -  \eta_R^{\ell} \frac{\partial {\cal L}_{CE}}{\partial \theta_R}\label{eq:train_b},  \\	
				\theta_F^{\ell+\frac{1}{2}} &  \leftarrow   \theta_F^{\ell}   -  \eta_F^{\ell} \frac{\partial {\cal L}_{CE}}{\partial \theta_F}	\label{eq:train_c}	,
				\vspace{-2pt}
			\end{align}
		\end{subequations} 	\vspace{-5pt}
		\item Update the feature extractor for fixed  $(\theta_E^{\ell+1}, \theta_R^{\ell+1})$:
		\begin{equation} \label{eq:train2}
			\theta_F^{\ell+1}    \leftarrow   \theta_F^{\ell+\frac{1}{2}}   -  {\eta}_F^{\ell}( \lambda_1 \frac{\partial {\mathcal L}_{MI}}{\partial \theta_F}	-   \lambda_2 \frac{\partial {\mathcal L}_{IE}}{\partial \theta_F})	,		
		\end{equation}
	\end{enumerate}   
	where $\ell$ is the iteration index and $\eta^\ell$ is the step-size for the $\ell$-th iteration. 
	
	We should mention that when updating the classifier, we have also performed an intermediate update for the feature extractor in~\eqref{eq:train_c}. This intermediate update is crucial to stabilize the learning process, since  $\theta_F^{\ell+\frac{1}{2}} $ ties up the update of the  classifier and the feature extractor.
	
	\section{Distributed RIEI}\label{section:FedRIEI}
	%In the last section, we have proposed an RFFI model under the premise that all the source domain data are collected and the RFFI model is trained centrally. However, in practice, the multiple source domain data may be distributively stored, and due to data privacy or  In this section, we will focus on the distributed training of the model.  

	The previous RIEI model has been developed under the premise of centralized training. In this section, we consider a more practical scenario, where all the receivers are distributed, and it is impossible for the remote receivers to transmit all the raw RF data to the server due to data privacy or bandwidth limitation. To address these challenges, in this section we leverage federated learning (FL) to train RIEI  in a distributed manner.

	\subsection{Federated Learning}
	
\begin{figure}[]
		\centering
		\centerline{\includegraphics[width=8.5cm]{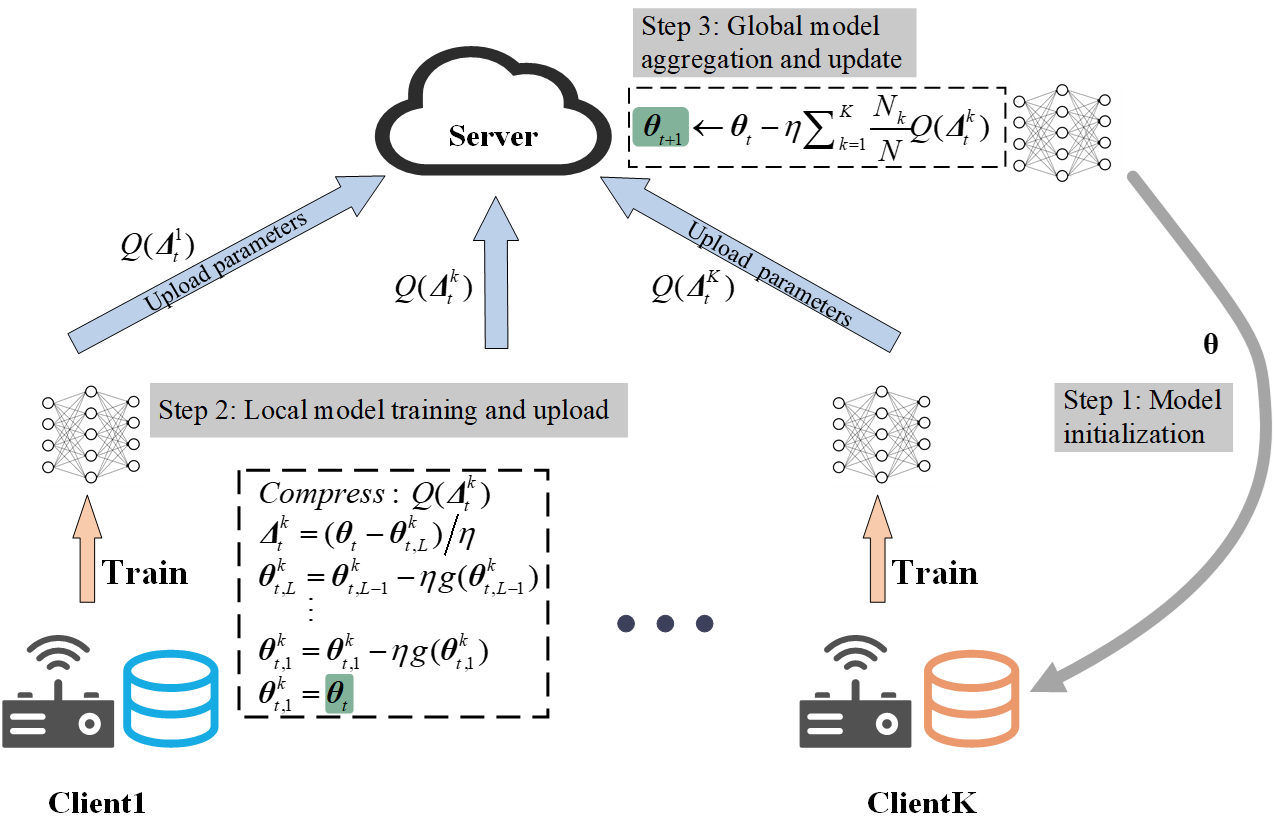}}
		%\vspace{1.5cm}
		\caption{The federated learning framework (with compression).}\label{fl}
		%\vspace{-5pt}
	\end{figure}

	  FL is a promising privacy-preserving distributed learning paradigm that protects privacy by learning a global model across multiple devices without exposing local datasets. In FL, each device is referred to as a client. Fig.~\ref{fl} illustrates the FL framework, which includes $K$ clients and a centralized server. The training of a global model in FL requires numerous iterations involving both the server and the clients. In each iteration, the server sends an initialized global model to all clients. Each client then trains this model on its local dataset. Subsequently, the model updates from each client are sent to the centralized server and are aggregated to produce a new global model. After several iterations, the global deep learning model is well trained. 
	
	The essence of this approach lies in keeping the original RF data localized, thus safeguarding privacy and reducing the need for centralized data storage. Instead, only model updates encapsulated as gradients are exchanged between local platforms and the central server. This decentralized learning strategy not only addresses the complexity of distributed RF data, but also allows for the flexible integration of data from various sources into the training process.

	\subsection{Federated RIEI (FedRIEI)}
	%Combining RIEI and FL, FedRIEI addresses the problem of learning across multiple data domains. By treating each receiver as a distinct domain,  learning receiver-agnostic RFFI model involves training the RFFI model on data from multiple receivers while encouraging the extraction of features that are independent of the specific receiver characteristics. To this end, we disentangle  the extracted feature into the receiver-dependent feature and the emitter-dependent feature via introducing the cross entropy loss, the mutual independence loss and the information entropy loss. By passing only the emitter-dependent feature to the subsequent  classification network, we attain a robust RFFI  model that is insensitive to the receiver characteristics and has good generalization ability to a new receiver. 
	
	%The detailed FedRIEI training is presented in Algorithm~\ref{method}. The FedRIEI training process includes two phases: server execution and client update.
	
	By fitting the RIEI network  into the FL framework, we can train the RIEI network in a distributed manner, and the resultant  network is coined as {\it FedRIEI}. The detail of FedRIEI training is given in  
	Algorithm~\ref{method}, which includes two phases: server execution and client update.

	\begin{algorithm}[t]
		\caption{\small FedRIEI Training Algorithm.}
		\label{method}
		\textbf{Input:} RF data ${\mathcal S} = \{ {\mathcal S}^k | k = 1 ,\ldots, K\}$, one-hot vector $\mathds{1}_m$,  model parameters of the FED, EC, RC $\boldsymbol{\theta} = \{ \theta_F,\theta_E, \theta_R \}$, total communication rounds $T$, number of local steps $L$, clients stepsize $\boldsymbol{\eta} = \{ \eta_F,\eta_E, \eta_R \}$.   
		
		\textbf{Output:} model parameters of the FEC, EC.
		\\
		
		\textbf{Server executes:}
		\begin{algorithmic}[1]
			\State initialize $\boldsymbol{\theta}_0$
			\For {round $t = 1, 2, \cdots, T$}
			\For {each client $k = 1, 2, \cdots, K$ \textbf{in parallel}}
			\State $\boldsymbol{\Delta}_t^k \gets$ ClientUpdate$(k, \boldsymbol{\theta}_t)$
			\EndFor
			\State $\boldsymbol{\theta}_{t+1} \gets \boldsymbol{\theta}_t -  {\eta}\sum_{k=1}^{K} \frac{N_k}{N} Q(\boldsymbol{\Delta}_t^k)$
			\EndFor        
		\end{algorithmic}
		
		\textbf{ClientUpdate}$(k, \boldsymbol{\theta}_t)$\textbf{:} // Run on client $k$
		\begin{algorithmic}[1]
			\State receive $\boldsymbol{\theta}_t = \{ \theta_F,\theta_E, \theta_R \}$ from server
			\For {each local epoch $\ell = 1, 2, \cdots, L$}            
			\State sample one mini-batch $S_{t,\ell}^k$ from $S^k$            
			\State update $\theta_E, \theta_R, \theta_F$ on $S_{t,\ell}^k$ to minimize ${\mathcal L}_{CE}$
			\State update $\theta_F$ on $S_{t,\ell}^k$ to minimize $\lambda_1 {\mathcal L}_{MI} - \lambda_2 {\mathcal L}_{IE}$
			\EndFor
			\State compress gradient $Q(\boldsymbol{\Delta}_t^k) = Q((\boldsymbol{\theta}_t - \boldsymbol{\theta}_{t,L}^k)/\boldsymbol{\eta}_t^k)$ based on SignSGD or 1-SignSGD or $\infty$-SignSGD
			\State upload the compressed gradient to server.
		\end{algorithmic} 
	\end{algorithm}
	
	{\bf Server Execution Phase.} The server is used to aggregate the model gradients uploaded by the clients. To start the training, the server initializes the network parameters $\theta = \{ \theta_F,\theta_E, \theta_R \}$ and distributes them to the clients. At each round, the server  collects the newly-learned gradients from all clients, federated averages them to obtain the new global parameters, and broadcast them back to each client at the beginning of next round. After $T$ rounds of server-client interaction, the parameters converge and the FedRIEI model is  well-trained. 
	
	{\bf Client Update Phase.} During the training process, the clients receive parameters $\theta$ from the server to train on the local data. Specifically, perform $L$ rounds of epoch training, where $L$ denotes the number of local training iterations  at each round. The minibatch data on client $k$ and communication round $t$ after $\ell$ local updates is expressed as $S_{t,\ell}^k$. As shown in Step 4, ${\mathcal L}_{CE}$ is used to control the training of EC, RC and FED. The parameters $\theta_E$, $\theta_R$, $\theta_F$ of EC, RC and FED are updated according to Eqn.~\eqref{eq:train}. In step 5, updating the parameters of FED  according to Eqn.~\eqref{eq:train2}. After local training is completed, the model gradients for FED, EC, and RC, denoted as $\boldsymbol{\Delta}$, are compressed by the quantizer  $Q$ (cf. step 7) and the quantized gradients are uploaded to the server.

	\begin{Remark}
		The compressed gradients, rather than the gradients,  are sent to the server because in practice, the data link between the client and the server  is usually rate-limited. After compression, the total number of transmitted bits during training can be effectively reduced. We will discuss different compression schemes, including SignSGD, 1-SignSGD and $\infty$-SignSGD, and compare their performance in Section~\ref{sec:experiment_compression}.
		%In this work, we focus on one-bit quantization and compare several one-bit compression schemes, including  SignSGD \cite{r43_bernstein2018signsgd}, $1-$SignSGD \cite{r41_tang2023z} and $\infty-$SignSGD \cite{r41_tang2023z}.
		%SignSGD a simple yet elegant technique is to take the sign of each coordinate of the local gradients, which requires only one bit for transmitting each coordinate. For any $x \in \mathbb{R}$, we define the sign operator as: $Sign(x) = 1$ if $x \geq 0$ and $-$$1$ otherwise. 	$1-$SignSGD, a method is employed that initially introduces noise, following a Gaussian distribution with a mean of $0$ and a variance of $\sigma$, to the local gradient. Subsequently, a sign operation is applied to each coordinate of the local gradients. Similarly, only one bit is necessary for transmitting each coordinate. $\infty-$SignSGD \cite{r41_tang2023z}, similar to $1-$SignSGD, except that the noise obeys a uniform distribution.  
	\end{Remark}

	\begin{Remark}	
		The convergence and stability of FedRIEI is guaranteed by using the convergence results in~\cite{r41_tang2023z}. Specifically, it is shown in \cite{r41_tang2023z} that when $\infty$-Sign  and $1$-Sign compression schemes are applied for client-uploaded gradients, the resultant $\infty$-SignSGD and $1$-SignSGD 			
		converge  to  stationary points (with vanishing gradients under $\ell_2$-norm measure) at  rates of  $\mathcal{O}(\tau^{-1/2})$ and $\mathcal{O}(\tau^{-1/3})$, resp.  Here, $\tau=TL$, where $T$ represents the number of communication rounds between the server and clients, and $L$ denotes the number of training iterations per round performed by each client.
		
	\end{Remark}

	\section{Experiments}\label{section:Experiments}
	In this section, we conduct experiments on the publicly available Wisig dataset \cite{r16_hanna2022wisig}  and our own HackRF dataset to evaluate the performance of RIEI under both centralized and distributed settings. 
	
	\subsection{Experimental Setup}
	1) Dataset: The Wisig dataset is constructed using WiFi nodes as emitters, emitting signals to another WiFi access point, and using multiple USRPs to receive WiFi signals. We pick a subset of the Wisig dataset with six transmitters (Tx) and six receivers (Rx). The subset is divided into six domains, each corresponding to one receiver’s data. The dataset structure is presented in Table \ref{wisigoverall}. As the receiver is positioned within a two-dimensional (2D) plane, the location of receivers is identified by their horizontal and vertical coordinates. For instance, the receivers situated at position $(1,1)$ in both horizontal and vertical planes are denoted as receiver $R{x_{(1 - 1)}}$. We perform cross-receiver tests, randomly selecting one receiver as target domain and using the other receivers as source domains. The well-trained model is tested on the target domain data. 
	
\begin{figure}%[htb] 
		\begin{minipage}[b]{0.48\linewidth}
			\centering
			\centerline{\includegraphics[width=5cm]{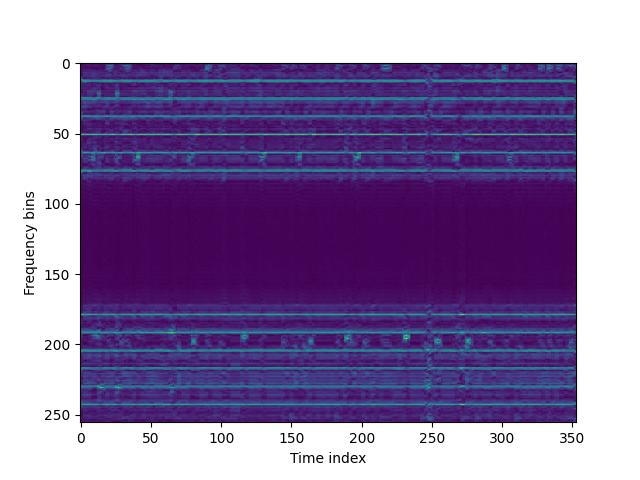}\label{wisig_sp}}
			%  \vspace{1.5cm}
			\centerline{(a) Wisig}\medskip
		\end{minipage}
		\hfill
		\begin{minipage}[b]{0.48\linewidth}
			\centering
			\centerline{\includegraphics[width=5cm]{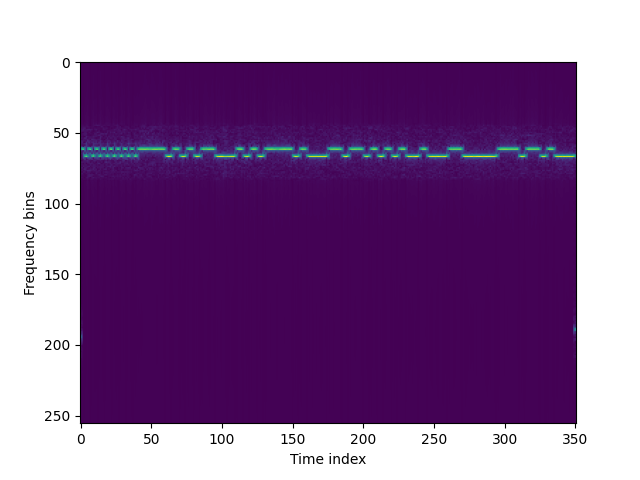}\label{HackRF_sp}}
			%  \vspace{1.5cm}
			\centerline{(b) HackRF}\medskip
		\end{minipage}
		\caption{The Spectrogram of signal.}
		\label{Spectrogram}     
	\end{figure}

	The HackRF dataset is created by our laboratory through the open-source HackRF platform. Four HackRFs serve as emitters, and three HackRFs as receivers. The collection process involves Matlab, with signals generated at a 2.5GHz frequency. Signals are modulated using BFSK at a 2MHz sampling rate. Amplitude and frequency noise are added uniformly to introduce randomness. Data is randomly partitioned for training and testing, similar to the Wisig dataset. The signal spectrum of the Wisig dataset and HackRF dataset are shown in Fig.~\ref{Spectrogram}. Table \ref{wisigoverall} provides an overview of the HackRF dataset structure.

%%	\begin{figure}%[htb] 
%		%
%		\begin{minipage}[b]{0.48\linewidth}
%			\centering
%			\centerline{\includegraphics[width=3.8cm]{fig/stft.png}\label{wisig_sp}}
%			%  \vspace{1.5cm}
%			\centerline{(a) Wisig}\medskip
%		\end{minipage}
%		\hfill
%		\begin{minipage}[b]{0.48\linewidth}
%			\centering
%			\centerline{\includegraphics[width=3.8cm]{fig/filter_stft_2.png}\label{HackRF_sp}}
%			%  \vspace{1.5cm}
%			\centerline{(b) HackRF}\medskip
%		\end{minipage}
%		%
%		\caption{The Spectrogram of signal.}
%		\label{Spectrogram}     
%	\end{figure} 

	\begin{table}[H] 
		\centering
		\caption{The overall structure of the Wisig and HackRF dataset.\label{wisigoverall}}
		%	\newcolumntype{C}{>{\centering\arraybackslash}X}
		\begin{tabular}{ccc}
			\hline
			\textbf{Parameter}	& \textbf{Wisig}	& \textbf{HackRF}\\
			\hline
			Emitter		& 6 & 4\\
			Receiver		& 6 & 3 \\
			Input sample dimension		& $2 \times 256$ & $2 \times 224 \times 224$\\
			Training samples per receiver		& 14400  & 3200\\
			Testing samples per receiver		& 4800 & 800 \\
			\hline
		\end{tabular}
		% \noindent{\footnotesize{\textsuperscript{1} Tables may have a footer.}}
	\end{table}

	\begin{table*}[ht]
		\renewcommand\arraystretch{1.3}
		\centering
		\caption{Comparison of Performance on the HackRF dataset.}\label{hacrf}
		\begin{tabular}{cccccccc}
			%			\toprule
			\hline
			\textbf{Input Data Form} & \textbf{Feature Extraction} &\multicolumn{1}{c}{\textbf{Model}}&
			\multicolumn{1}{c}{\textbf{FLOPs(M)}} &
			\multicolumn{1}{c}{\textbf{Params(M))}} & 
			\multicolumn{1}{c}{\textbf{HackRF $1$ Acc(\%)}} & \multicolumn{1}{c}{\textbf{HackRF $2$ Acc(\%)}} & \multicolumn{1}{c}{\textbf{HackRF $3$ Acc(\%)}}  \\ \hline
			\multirow{4}*{ \makecell[c]{One-dim. \\ temporal signal}} &
			\multirow{4}*{ \makecell[c]{ResNet$1$D-$18$}} 
			& Baseline  & $0.799$ &    $0.022$ & $57.73 \pm3.37$ &    $68.40 \pm 3.65$ &   $49.98 \pm 0.12$\\ 
			&  & DANN & $0.866$ &    $0.088$ & $61.92 \pm 5.11$ &    $74.08 \pm 3.37$ &   $52.30 \pm 2.54$\\
			&  & SD-RXA  &$0.933$ & $0.156$ & $63.38 \pm 0.78$ &  $72.00 \pm 0.75$ &   $62.55 \pm 0.32$\\
			&   & RIEI &$0.867$  & $0.089$   & \textbf{69.92 $\pm$ 4.55}  & \textbf{74.48 $\pm$ 5.48} & \textbf{64.22 $\pm$ 4.76}\\ \hline
			\multirow{4}*{\makecell[c]{Two-dim. \\ time-frequency \\ diagram}} &
			\multirow{4}*{\makecell[c]{ResNet$50$}}
			& Baseline  & $4095$ &    $26.13$ & $34.56 \pm 2.44$ &    $54.81 \pm 2.44$ &   $54.24 \pm 6.76$\\
			& & DANN  & $4095$ &    $26.19$ & \textbf{55.15 $\pm$ 1.46} &    $75.68 \pm 3.85$ &   $68.55 \pm 0.65$\\
			& & SD-RXA & $4095$ & $26.26$   & $52.73 \pm 0.73$ &   $76.03 \pm 0.86$ &   $70.35 \pm 0.46$\\
			& & RIEI   &$4095$  & $26.20$ &$53.41 \pm 5.21$  & \textbf{79.21 $\pm$ 6.79} & \textbf{73.22 $\pm$ 2.60}\\ \hline
			
		\end{tabular}
	\end{table*}
	
	2) Model setup: The FED employs Residual Networks (ResNets) \cite{r42_he2016deep} for both the Wisig and HackRF datasets. Both one-dimensional temporal data and two-dimensional time-frequency spectrogram data   are considered as input. Specifically, ResNet$1$D-$18$ and ResNet$50$ with attention are utilized as FED modules   when the input data is in the  two-dimensional time-frequency spectrogram form. When the input data is in the one-dimensional temporal form, we replace the 2D-convolution with 1D-convolution in ResNet to better accommodate the one-dimensional time series of RF signals. Both the EC and RC networks are implemented with three-layer fully connected networks. The hyperparameters including the initial learning rate $\eta$ of the FED, EC, RC are all 0.0001. Additionally, both $\lambda_1$ and $\lambda_2$ in Eq.~\eqref{eq:overall_loss} are set to 1.2 to optimize the model performance. 
	
	% \subsection{Simulation Results}
	% In the simulation experiment, we first demonstrate the advantages of RIEI by comparing it with the baseline approach under a centralized scenario. Then, we increase the number of clients and compare FedRIEI with other solutions. Finally, with a focus on reducing communication costs, we introduce gradient compression into the FedRIEI framework.
	
	\subsection{Performance of Centralized Scenario}
	%In the simulation experiment, we first demonstrate the advantages of RIEI by comparing it with the baseline approach. Then, we have an ablation study of the RIEI scheme. Finally, we observe the disentangled emitter-related features and receiver-related features through plotted reduced-dimensional scatter plots.
	
	\subsubsection{Comparison with Existing Methods} 
	We first carry out the classification experiments on the centralized network to demonstrate the benefits of RIEI by comparing it with the baseline method  \cite{al2020exposing}, the receiver-agnostic RFFI method based on Domain-Adversarial Neural Networks (DANN)~\cite{r14_shen2023towards},  and the statistical distancebased receiver-agnostic (SD-RXA) method~\cite{zhao2024gan}. The baseline method  treats all the multi-source data as a whole, and  trains an RFFI model  by minimizing the  cross-entropy loss. The DANN method aims at learning the feature representation while reducing the distributional differences between the source and target domains through adversarial training. 
	The SD-RXA optimizes the feature extractor to ensure that the extracted receiver-related and emitter-related features tend to come from distributions with significant statistical distances. In the Wisig and HackRF experiments, we feed the models with time-domain signals and time-frequency diagrams obtained through Short-time Fourier Transform (STFT). The experimental design entails training the model on data from two receivers, followed by testing on data from a third receiver. To evaluate the model, we take into account the mean and standard deviation of the classification accuracy for the last 5 epochs. 
	
	In Table \ref{hacrf}, we present the computational complexity, model parameter counts, and the classification accuracy of RIEI, SD-RXA, DANN, and the baseline method on the HackRF dataset. The data in the HackRF $i$ column were tested using data from the HackRF $i$ receiver and trained using data from the remaining two receivers. For computational complexity, we use FLOPs (floating point operations per second) as the metric. It can be seen that the difference in computational complexity between RIEI, SD-RXA, DANN, and the baseline method is not significant. As for the recognition accuracy, the results clearly show that RIEI outperforms the other three methods. Particularly, when the inputs are time-domain signals, RIEI achieves an average improvement of $10.84\%$ over the baseline, $6.77\%$ over DANN, and $3.56\%$ over SD-RXA. When the inputs are time-frequency diagrams, it improves by an average of $20.74\%$ compared to the baseline, $2.15\%$ compared to DANN, and $2.24\%$ compared to SD-RXA.

	%	  The results clearly depict a penchant for RIEI to outshine the baseline in terms of recognition accuracy, with an average improvement of 20.74\%. 
	
	Table \ref{wisigacc_TD} and Table  \ref{wisigacc_FD} show the classification accuracy for the tasks on the Wisig dataset. Again, RIEI outperforms the other   three methods. Particularly, when using time-domain signals as inputs, RIEI achieves an average improvement of $9.37\%$, $8.38\%$ and $5.83\%$ over the baseline, DANN and SD-RXA methods, resp. With time-frequency diagrams as inputs, it improves by an average of $9.49\%$, $6.28\%$ and $4.27\%$ compared to the baseline, DANN and SD-RXA, resp.
		
	Based on the above results, it can be seen that the overall recognition performance is better when  one-dimensional temporal signals are used as input. This may be due to  electromagnetic signals are temporal sequences, and one-dimensional feature extraction networks can effectively capture local characteristics and temporal dependencies within the time series.

	\begin{table*}[h!]
		\renewcommand\arraystretch{1.3}
		\centering
		\caption{Comparison of recognition accuracy (\%) on the Wisig dataset (Time-domain signal).}\label{wisigacc_TD}
		\scalebox{0.9}{
			\begin{tabular}{cccccc}
				%			\toprule
				\hline
				\textbf{Training receivers}& \textbf{Test receiver}& \textbf{Baseline}& \textbf{DANN}& \textbf{SD-RXA}& \textbf{RIEI}\\
				\hline
				$Rx_{(1-1)}$, $Rx_{(7-7)}$  & $Rx_{(1-19)}$   & $70.39 \pm 1.37$&  $72.15 \pm 0.36$&  $73.70 \pm 0.33$   &  \textbf{77.88 $\pm$ 2.23}   \\
				$Rx_{(1-1)}$, $Rx_{(8-8)}$  & $Rx_{(1-19)}$   & $62.85 \pm 0.75$&  $62.52 \pm 0.32$&  $70.53 \pm 0.88$   &  \textbf{79.43 $\pm$ 1.66}   \\
				$Rx_{(1-1)}$, $Rx_{(14-7)}$  & $Rx_{(1-19)}$   & $64.78 \pm 0.72$&  $63.43 \pm 0.11$&  $64.63 \pm 0.35$   &  \textbf{66.09 $\pm$ 0.67}    \\
				$Rx_{(7-7)}$, $Rx_{(8-8)}$  & $Rx_{(1-19)}$   & $69.16 \pm 1.26$&  \textbf{71.72 $\pm$ 1.18}&  $68.04 \pm 1.20$   &  $70.51 \pm 3.53$    \\
				$Rx_{(7-7)}$, $Rx_{(14-7)}$  & $Rx_{(1-19)}$   & $65.34 \pm 0.54$&  $64.89 \pm 0.19$&  $68.10 \pm 0.68$   &  \textbf{77.35 $\pm$ 1.53}    \\
				$Rx_{(8-8)}$, $Rx_{(14-7)}$  & $Rx_{(1-19)}$   & $62.51 \pm 0.80$&  $63.67 \pm 0.34$&  $65.16 \pm 0.13$   &  \textbf{75.48 $\pm$ 1.21}  \\
				\hline
				$Rx_{(1-1)}$, $Rx_{(1-19)}$  & $Rx_{(14-7)}$   & $66.54 \pm 1.31$&  $70.75 \pm 0.61$ &  $71.69 \pm 0.25$  &  \textbf{71.91 $\pm$ 2.08}    \\
				$Rx_{(1-1)}$, $Rx_{(7-7)}$  & $Rx_{(14-7)}$   & $60.49 \pm 1.57$&  $58.18 \pm 0.86$&  $66.40 \pm 0.49$   &  \textbf{68.33 $\pm$ 2.37}    \\
				$Rx_{(1-1)}$, $Rx_{(8-8)}$  & $Rx_{(14-7)}$   & $61.10 \pm 0.56$&  $63.18 \pm 0.46$&  $64.84 \pm 0.26$   &  \textbf{73.54 $\pm$ 1.27}   \\
				$Rx_{(1-19)}$, $Rx_{(7-7)}$  & $Rx_{(14-7)}$   & $62.97 \pm 1.89$&  $61.65 \pm 0.55$&  $63.20 \pm 0.31$   &  \textbf{73.52 $\pm$ 3.15}    \\
				$Rx_{(1-19)}$, $Rx_{(8-8)}$  & $Rx_{(14-7)}$   & $65.61 \pm 2.41$&  $66.51 \pm 0.20$&  $68.40 \pm 0.19$   &  \textbf{72.05 $\pm$ 2.71}    \\
				$Rx_{(7-7)}$, $Rx_{(8-8)}$  & $Rx_{(14-7)}$   & $55.31 \pm 1.48$&  $60.23 \pm 0.25$&  $64.85 \pm 1.07$   &  \textbf{73.46 $\pm$ 2.00}  \\
				\hline
			\end{tabular}
		}
	\end{table*}

	\begin{table*}[h!]
		\renewcommand\arraystretch{1.2}
		\centering
		\caption{Comparison of recognition accuracy (\%) on the Wisig dataset (Time-frequency diagram).}\label{wisigacc_FD}
		\scalebox{0.9}{
			\begin{tabular}{cccccc}
				%			\toprule
				\hline
				\textbf{Training receivers}& \textbf{Test receiver}& \textbf{Baseline}& \textbf{DANN}& \textbf{SD-RXA}&  \textbf{RIEI}\\
				\hline
				$Rx_{(1-1)}$, $Rx_{(7-7)}$  & $Rx_{(1-19)}$   & $68.56 \pm 0.55$&  $68.71 \pm 0.53$&  $69.44 \pm 0.34$   &  \textbf{70.71 $\pm$ 0.89}   \\
				$Rx_{(1-1)}$, $Rx_{(8-8)}$  & $Rx_{(1-19)}$   & $56.69 \pm 0.30$&  $66.71 \pm 0.83$&  $67.34 \pm 0.66$   &  \textbf{73.70 $\pm$ 0.92}   \\
				$Rx_{(1-1)}$, $Rx_{(14-7)}$  & $Rx_{(1-19)}$  & $55.68 \pm 0.80$&  $59.49\pm 0.81$&  $64.00 \pm 1.09$   &  \textbf{64.92 $\pm$ 0.41}    \\
				$Rx_{(7-7)}$, $Rx_{(8-8)}$  & $Rx_{(1-19)}$   & $62.16\pm 0.53$&  \textbf{68.08 $\pm$ 0.50}&  $67.67 \pm 1.24$   &  $64.75 \pm 0.88$    \\
				$Rx_{(7-7)}$, $Rx_{(14-7)}$  & $Rx_{(1-19)}$   & $66.90\pm 1.05$&  $67.23 \pm 0.83$&  $67.17 \pm 0.47$   &  \textbf{74.70 $\pm$ 1.65}    \\
				$Rx_{(8-8)}$, $Rx_{(14-7)}$  & $Rx_{(1-19)}$  & $63.87 \pm 0.31$&  $63.99 \pm 0.29$&  $63.56 \pm 0.46$   &  \textbf{68.31 $\pm$ 0.57}  \\
				\hline
				$Rx_{(1-1)}$, $Rx_{(1-19)}$  & $Rx_{(14-7)}$  & $57.77 \pm 1.21$&  $58.73 \pm 0.89$&  $60.90 \pm 0.86$   &  \textbf{77.69 $\pm$ 1.69}    \\
				$Rx_{(1-1)}$, $Rx_{(7-7)}$  & $Rx_{(14-7)}$   & $58.61 \pm 0.21$&  $65.31 \pm 0.96$&  $70.08 \pm 0.85$   &  \textbf{73.79 $\pm$ 1.10}    \\
				$Rx_{(1-1)}$, $Rx_{(8-8)}$  & $Rx_{(14-7)}$   & $44.42 \pm 1.15$&  $49.11 \pm 0.76$&  $53.87 \pm 0.80$   &  \textbf{52.71 $\pm$ 0.64}   \\
				$Rx_{(1-19)}$, $Rx_{(7-7)}$  & $Rx_{(14-7)}$  & $45.25 \pm 0.76$&  $49.94 \pm 1.39$&  $45.19 \pm 0.87$   &  \textbf{52.05 $\pm$ 0.89}    \\
				$Rx_{(1-19)}$, $Rx_{(8-8)}$  & $Rx_{(14-7)}$  & $52.26 \pm 0.32$&  $51.40 \pm 0.50$&  $58.76 \pm 1.40$   &  \textbf{60.11 $\pm$ 1.04}    \\
				$Rx_{(7-7)}$, $Rx_{(8-8)}$  & $Rx_{(14-7)}$   & $56.68 \pm 0.72$&  $58.58 \pm 0.90$&  $61.32 \pm 1.15$  &  \textbf{67.10 $\pm$ 1.32}  \\
				\hline
			\end{tabular}
		}
	\end{table*}

	\subsubsection{Ablation Studies}
	Ablation studies are conducted to demonstrate the efficacy of various components in RIEI. As presented in Table \ref{efusion}, the notation (H$i$, H$j$)$ \to $H$k$ indicates that the model is trained using data from  HackRF receivers $i$ and $j$, and tested with data from  HackRF receiver $k$, while ($Rx_{(x-y)}$,\,$Rx_{(x'-y')}$) $ \to $ $Rx_{(x''-y'')}$ signifies an identical process. Results indicate that using either information entropy loss or mutual independence loss alone results in an improved recognition accuracy of 14.79\% and 14.09\%, resp. Combining both loss functions results in  recognition accuracy improvement of 18.16\% compared to the baseline model. Furthermore, using both loss functions has better results compared to using only one of them, with accuracy  improved by 3.38\% and 4.08\%, resp. It is therefore concluded that both information entropy loss and mutual independence loss are effective for optimizing network performance.
	
	\begin{table*}[ht]
		\caption{Comparison of recognition accuracy (\%) of ablation studies.}\label{efusion} 	\vspace{-5pt}
		\renewcommand\arraystretch{1.3}
		\centering
		\begin{tabular}{c c c c c c c c}
			\hline
			\multicolumn{2}{c}{\textbf{Components}} & (H1,\,H2)  & (H1,\,H3)  & (H2,\,H3)  & ($Rx_{(1-1)}$, $Rx_{(8-8)}$) & ($Rx_{(7-7)}$, $Rx_{(14-7)}$) & ($Rx_{(7-7)}$, $Rx_{(8-8)}$)\\
			\cline{1-2}
			\textbf{IE} &\textbf{MI} & $\rightarrow$ H3 & $\rightarrow$ H2 & $\rightarrow$ H1 &  $\rightarrow$ $Rx_{(1-19)}$ & $\rightarrow$ $Rx_{(1-19)}$ & $\rightarrow$ $Rx_{(14-7)}$ \\
			\hline
			$\times$ & $\times$  &$54.24 \pm 6.76$  &    $54.81 \pm 2.44$  &   $34.56 \pm 2.44$  &  $62.85 \pm 0.75$   &   $65.34 \pm 0.54$&  $55.31 \pm 1.48$\\
			\hline
			\checkmark & $\times$ & $67.30 \pm 14.17$  &  $72.14 \pm 10.14$ & $52.77 \pm 6.98$  & $76.94 \pm 0.86$ &  $74.53 \pm 0.76$ &   $72.15 \pm 1.23$  \\
			$\times$ & \checkmark   & $72.40 \pm 4.10$  &$72.90 \pm 4.90$ &$52.25 \pm 4.63$ & $77.74 \pm 1.41$   & $68.12 \pm 2.32$     &$68.21 \pm 1.56$   \\
			\hline
			\checkmark & \checkmark &  $\bm{73.22 \pm 2.60}$    & $\bm{79.21 \pm 6.79}$  & $\bm{53.41 \pm 5.21}$ & $\bm{79.43 \pm 1.66}$  & $\bm{77.35 \pm 1.53}$ &  $\bm{73.46 \pm 2.00}$  \\

			\hline
		\end{tabular}
		%\vspace{-15pt}
	\end{table*}
	
	\subsubsection{t-SNE Visualization}
\begin{figure} \centering
		\subfigure[Emitter-dependent feature on H$1$.] {
			\label{h2}
			\includegraphics[width=0.45\columnwidth]{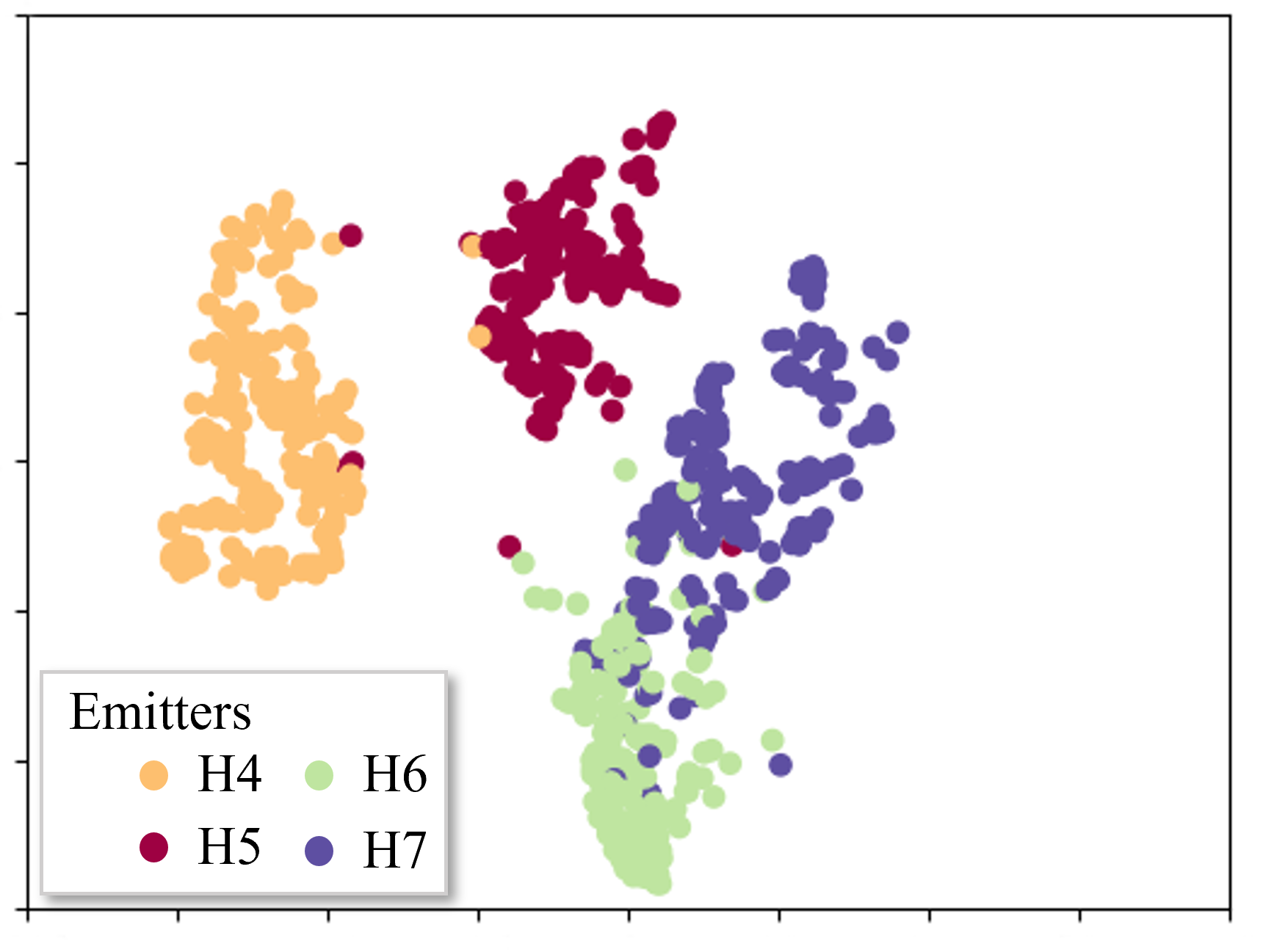}
		}
		\subfigure[Emitter-dependent feature on H$2$.] {
			\label{h4}
			\includegraphics[width=0.45\columnwidth]{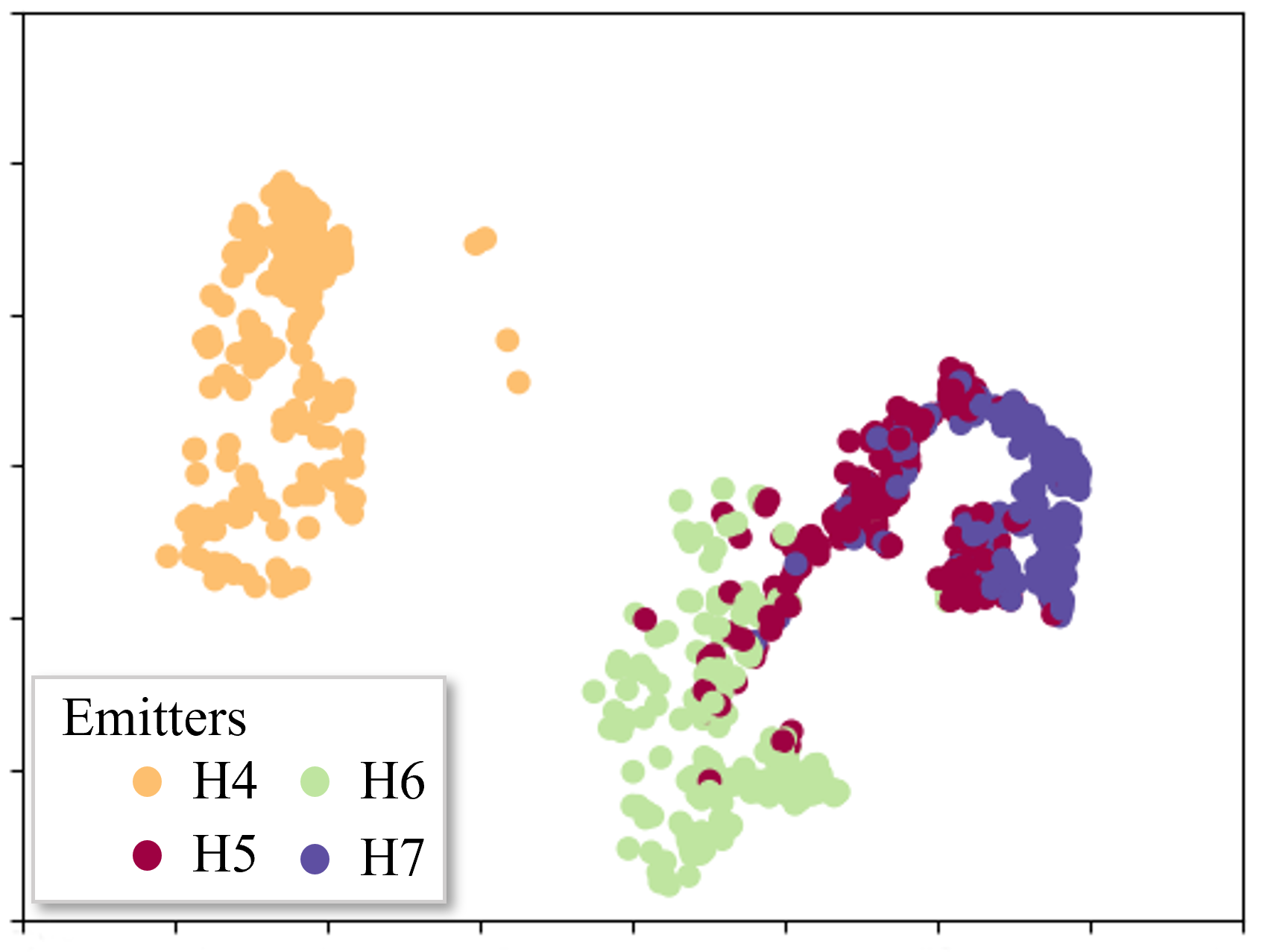}
		}
		\subfigure[Emitter-dependent feature on H$3$.] {
			\label{h3}
			\includegraphics[width=0.45\columnwidth]{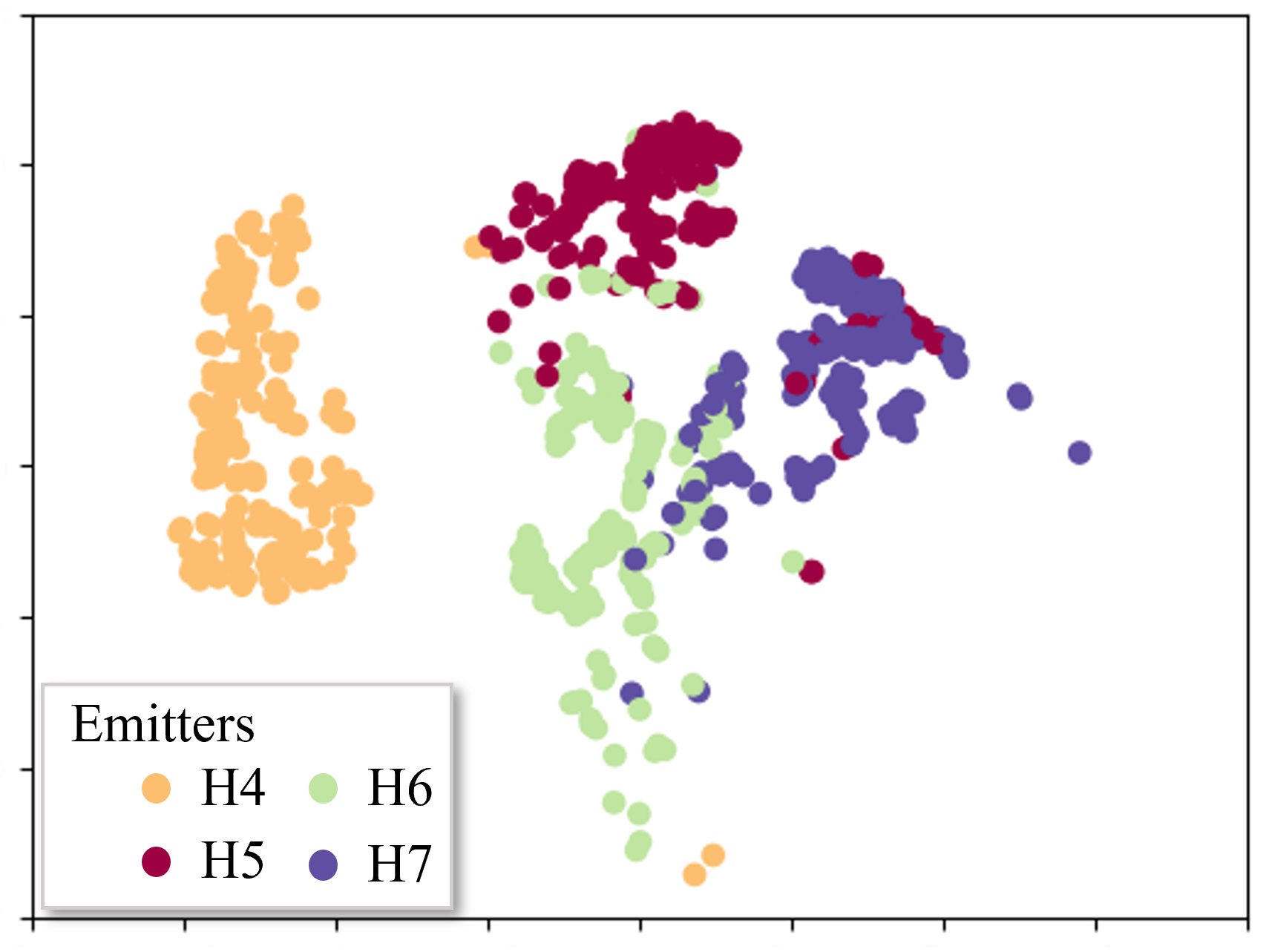}
		}
		\subfigure[Feature of Baseline  on H$3$.] {
			\label{tradition}
			\includegraphics[width=0.45\columnwidth]{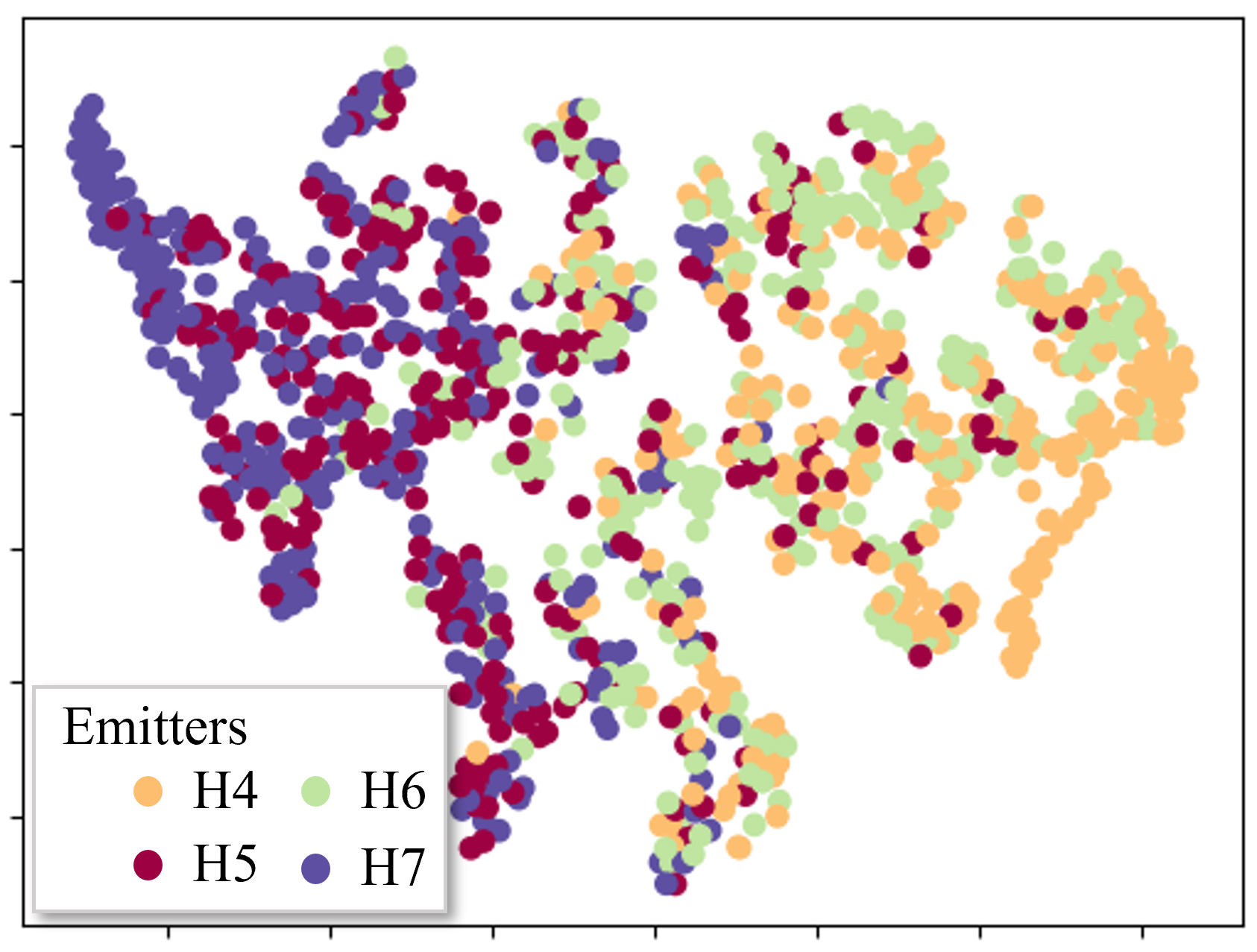}
		}	
		\caption{Visualization with t-SNE of emitter-dependent features. Different colors represent the features of different emitters.}
		\label{fig:tsneh}
		% \vspace{-10pt}
	\end{figure}

	In this example, we show case that the proposed RIEI model indeed disentangles the features of emitters and receivers by drawing  the scatter plot of the dimension-reduced emitter-dependent features. For illustration, we use  HackRF 1 and HackRF 2 to train the model and test it on  HackRF 3. Fig.~\ref{h2}-Fig.~\ref{h3} show the emitter-dependent features obtained by t-distributed Stochastic Neighbor Embedding (t-SNE) with different colors representing different emitters. As seen, the emitter-dependent features of    HackRF 3 follow a similar distribution as HackRF 1 and  HackRF 2, which implies that the proposed model can effectively mitigate the effect of receivers and make the feature distributed  consistently over different receivers. By contrast, Fig.~\ref{tradition} presents the   t-SNE visualization  obtained by the baseline method. Clearly, different emitters' features are densely overlapped due to the ignorance of the receiver effect. 
	
	\subsubsection{Robustness Test}
	In a wireless communication environment, the signal may be subject to various kinds of interference. In order to verify the robustness of the model, we consider adding two types of disturbances to the original signal: narrowband hopping  interference and broadband Gaussian noise. Narrowband hopping interference randomly and uniformly selects part of the frequency band to interfere. In our simulation, the whole signal frequency band is divided into  $256$ frequency bins and  $2$ of them are selected. The selection is  consistent within each data sample but varies randomly between samples. Broadband Gaussian noise  adds Gaussian noise across the entire frequency band of the signal.
	
\begin{figure}[]
		\centering
		\centerline{\includegraphics[width=9cm]{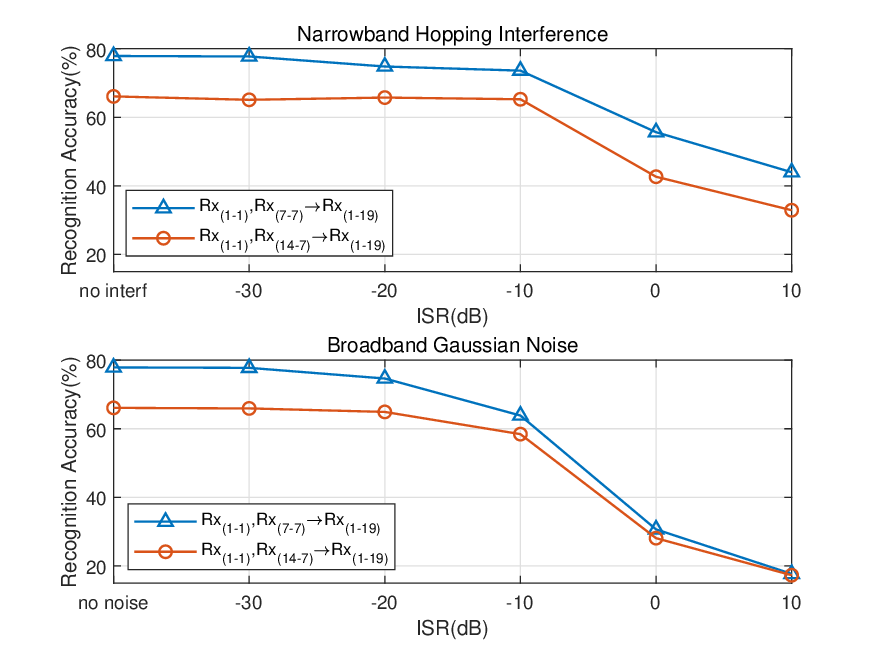}}
		\caption{The recognition accuracy of RIEI under different ISR.}\label{interference}
	\end{figure}

	Fig.~\ref{interference}  shows the recognition performance of RIEI under different interference-to-signal power ratios (ISRs).  From the figure, we see that the recognition accuracy of RIEI decreases as the ISR increases. Nevertheless, for a wide range of the ISR the recognition performance is quite stable, which implies that the proposed model RIEI model is not very sensitive to the disturbances. Moreover,  it can also be observed that the narrowband interference has less impact on recognition performance compared to the broadband noise. This may be due to the narrowband interference only affects a small portion of the signal spectrum, leaving most of the frequency components intact, thereby causing relatively less disruption to the overall features and information of the signal.

	\begin{table*}[ht]
		\renewcommand\arraystretch{1.3}
		\centering
		\caption{Comparison of recognition accuracy (\%) for four receivers training and another receiver testing.\label{Fedacc}}
		%	\begin{adjustwidth}{-\extralength}{0cm}
			%		\newcolumntype{C}{>{\centering\arraybackslash}X}
			\begin{tabular}{ccccc}
				\hline
				\textbf{Training data receivers}& \textbf{Test data receiver}& \textbf{FedBase accuracy}& \textbf{Centralized RIEI accuracy}& \textbf{FedRIEI accuracy}\\
				\hline
				$Rx_{(1-1)}$, $Rx_{(1-19)}$, $Rx_{(7-7)}$, $Rx_{(8-8)}$  & $Rx_{(14-7)}$   & $62.11 \pm 0.09$   &  $73.31 \pm 2.22$&  $76.54 \pm 0.26$  \\
				%			\hline
				$Rx_{(1-1)}$, $Rx_{(1-19)}$, $Rx_{(7-7)}$,$Rx_{(14-7)}$  & $Rx_{(8-8)}$   & $66.35 \pm 0.46$   &  $67.76 \pm 3.63$&  $69.83 \pm 0.88$  \\
				%			\hline
				$Rx_{(1-1)}$, $Rx_{(1-19)}$, $Rx_{(8-8)}$, $Rx_{(14-7)}$  & $Rx_{(7-7)}$   & $62.39 \pm 0.65$   &  $79.04\pm 1.36$&  $75.07 \pm 0.85$  \\
				%			\hline
				$Rx_{(1-1)}$, $Rx_{(7-7)}$, $Rx_{(8-8)}$, $Rx_{(14-7)}$  & $Rx_{(1-19)}$  & $60.18 \pm 0.95$   &  $68.21 \pm 1.02$&  $75.32 \pm 1.26$  \\
				%			\hline
				$Rx_{(1-19)}$, $Rx_{(7-7)}$, $Rx_{(8-8)}$, $Rx_{(14-7)}$  & $Rx_{(1-1)}$   & $64.87 \pm 1.03$   &  $76.29 \pm 0.34$&  $75.32 \pm 0.55$  \\
				\hline
			\end{tabular}
			%	\end{adjustwidth}
		% \noindent{\footnotesize{* Tables may have a footer.}}
	\end{table*}

	% \begin{table}[H]
		% 	\caption{Comparison of recognition accuracy (\%) of ablation studies.}\label{efusion} 	
		% 	\renewcommand\arraystretch{1.3}
		% 	\centering
		% 	\begin{tabular}{c c c c}
			% 		\toprule
			% 		\multicolumn{2}{c}{\textbf{Components}}  & ($Rx_{(1-1)}$, $Rx_{(7-7)}$, $Rx_{(8-8)}$, $Rx_{(14-7)}$) & ($Rx_{(1-1)}$, $Rx_{(1-19)}$, $Rx_{(7-7)}$, $Rx_{(8-8)}$) \\
			% 		\cline{1-2}
			% 		\textbf{IE} & \textbf{MI} &  $\rightarrow$ $Rx_{(1-19)}$ & $\rightarrow$ $Rx_{(14-7)}$ \\
			% 		\hline
			% 		$\times$ & $\times$  &$60.18 \pm 0.95$  &    $62.11 \pm 0.09$ \\
			% 		\hline
			% 		\checkmark & $\times$ & $71.05 \pm 1.40$  &  $76.10 \pm 0.26$ \\
			% 		$\times$ & \checkmark   & $66.25 \pm 0.51$  &$73.86 \pm 0.48$ \\
			% 		\hline
			% 		\checkmark & \checkmark &  $75.32 \pm 0.55$    & $76.54 \pm 0.26$  \\		
			% 		\bottomrule
			% 	\end{tabular}
		% \end{table}

	\subsection{Performance of Distributed Scenario}

	\subsubsection{Comparison with Baseline}
	We apply the baseline method mentioned in the previous subsection to the distributed scenario as the federated baseline network, referred to as FedBase. In the distributed setting, we compare FedRIEI with the centralized RIEI and FedBase. For simplicity, in this subsection all the federated  models are trained without gradient compression, and the performance comparison with compression will be discussed in the next subsection.

	Assuming there are four clients in training phase, each using the dataset of one receiver. At each iteration, each client first downloads the global model from the server; then it trains locally once to obtain the local gradient; at last, each client uploads its local gradient to the central server, where a new global model is obtained. The above process repeats until some convergence criterion is met. In test phase, the well-trained model is tested on the dataset from the fifth receiver. Therefore, for FedRIEI, it differs from the  RIEI model only in the training stage, where some additional message exchange and parameter aggregation are needed. However, for the inference stage,  FedRIEI and RIEI have the same complexity.

	In Table \ref{Fedacc}, we present the classification accuracy achieved by FedBase, RIEI, and FedRIEI approaches. The results clearly depict a penchant for FedRIEI to outperform the FedBase in terms of recognition accuracy, with an average improvement of 11.24\%. Compared to centralized RIEI, FedRIEI demonstrates superior performance. This is probably attributed to the utilization of Batch Normalization (BN) layers, which play the role of preventing gradient vanishing/exploding and accelerating network training. For centralized RIEI, the BN utilizes all the data in a   minibatch of size 64 to calculate a single scaling factor for different receivers' data within  the minibatch.  By contrast, the BN of FedRIEI is conducted locally at each client  with a smaller minibatch size of 16, and  different normalization factors are utilized by  clients. Therefore, the  BN of FedRIEI has more degrees of freedom and  more fine-controlled minibatch size than that of centralized RIEI, which help train a better generalizable  RFFI model on the unseen target domain.

	\subsubsection{Comparison with Gradient Compression} \label{sec:experiment_compression}
	In this subsection, we investigate  the impact of gradient compression on the performance of FedRIEI. We focus on one-bit quantization and compare several one-bit compression schemes, including  SignSGD \cite{r43_bernstein2018signsgd}, $1-$SignSGD \cite{r41_tang2023z} and $\infty-$SignSGD \cite{r41_tang2023z}.
	SignSGD is a scheme that directly takes the sign of each coordinate of the gradient. For any $x \in \mathbb{R}$, the sign operator is defined as: $Sign(x) = 1$ if $x \geq 0$ and $-$$1$ otherwise; $1-$SignSGD intentionally adds Gaussian noise $\xi \sim {\cal CN}(0, \sigma^2) $ to the gradient before performing sign quantization, i.e. $Sign(x + \xi)$; $\infty-$SignSGD is similar to $1-$SignSGD, except that the noise follows  uniform distribution.
	
	%\begin{flushright}
	%	
	%\end{flushright} 
	%SignSGD a simple yet elegant technique is to take the sign of each coordinate of the local gradients, which requires only one bit for transmitting each coordinate.	$1-$SignSGD, a method is employed that initially introduces noise, following a Gaussian distribution with a mean of $0$ and a variance of $\sigma$, to the local gradient. Subsequently, a sign operation is applied to each coordinate of the local gradients. Similarly, only one bit is necessary for transmitting each coordinate. $\infty-$SignSGD \cite{r41_tang2023z}, similar to $1-$SignSGD, except that the noise obeys a uniform distribution.
	
\begin{figure}[]
		\centering
		\centerline{\includegraphics[width=8.5cm]{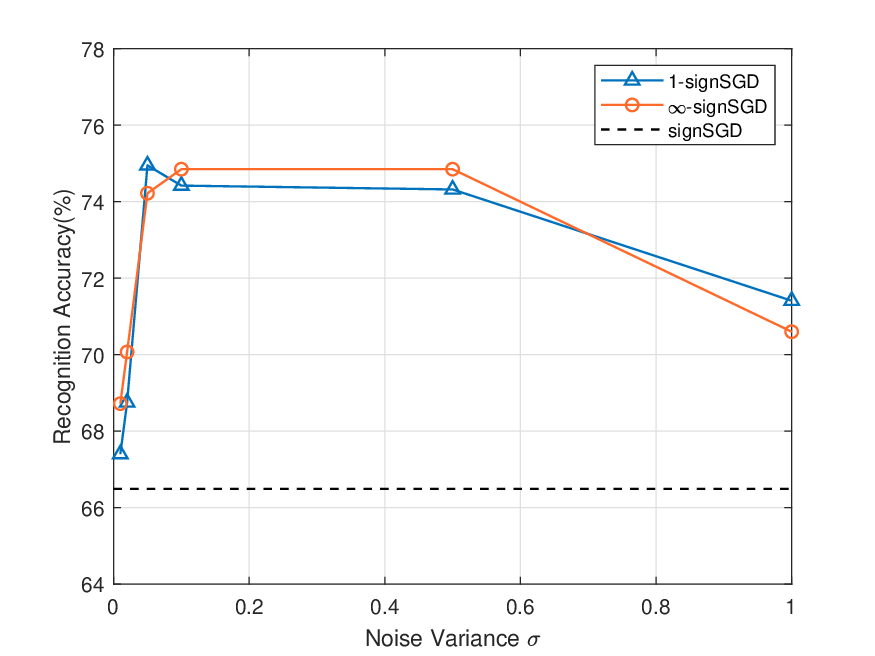}}
		\caption{The recognition accuracy of different gradient compression methods when generalizing to the receiver $Rx_{(1-19)}$ with different noise variances.}\label{1-19}
	\end{figure}

\begin{figure} \centering
		\subfigure[Test Loss v.s. bits] {
			\label{loss_bit}
			\includegraphics[width=0.9\columnwidth]{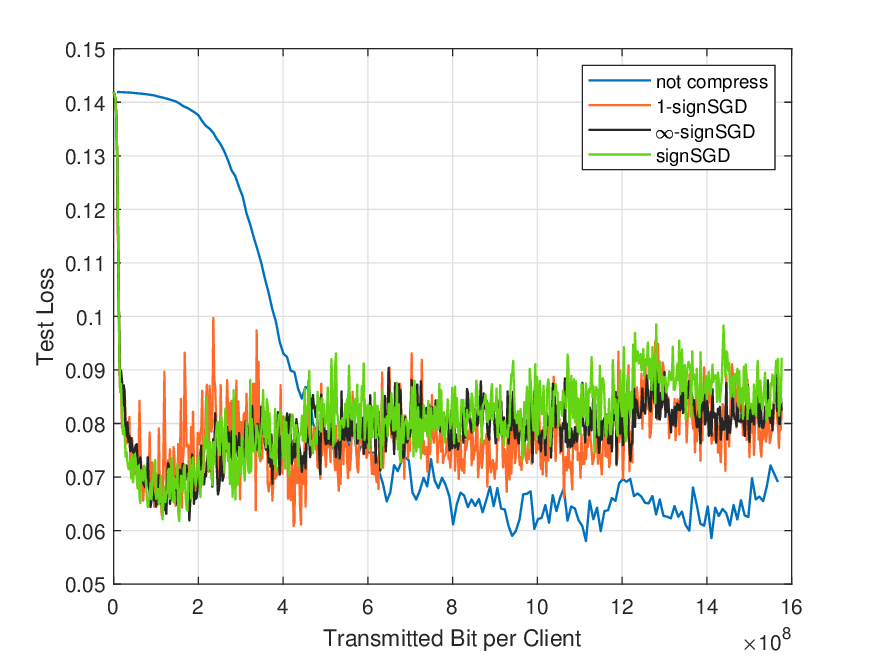}
		}
		\subfigure[Test Accuracy v.s. bits] {
			\label{acc_bit}
			\includegraphics[width=0.9\columnwidth]{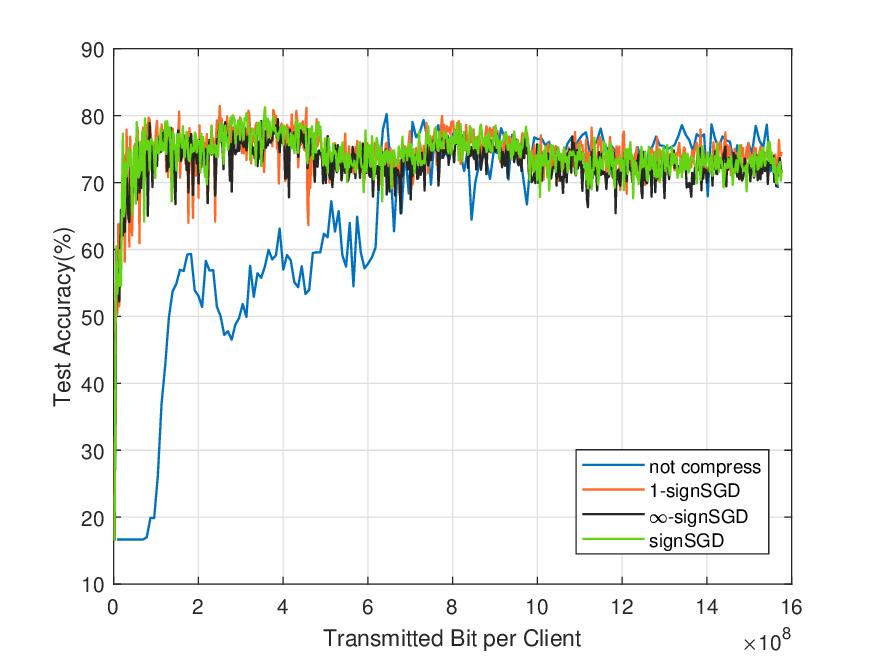}
		}
		
		\caption{Performance of FedRIEI with different gradient compression methods when generalizing ($Rx_{(1-1)}$, $Rx_{(1-19)}$, $Rx_{(7-7)}$, $Rx_{(8-8)}$)  to $Rx_{(14-7)}$.}
		\label{cost}
	\end{figure}

	Table \ref{compress} presents the performance of various gradient compression methods. It can be observed that the performance loss is minimal when compressing the gradients before uploading them to the server, with more pronounced losses only apparent when generalizing to the receiver $Rx_{(1-19)}$. Additionally, the approach of first adding noise to local gradients before applying the sign operation outperforms the direct signing of individual elements in the gradient. For the generalization to the receiver $Rx_{(1-19)}$, we adjusted the noise variance of $1-$SignSGD and $\infty-$SignSGD. Fig.~\ref{1-19} shows the performance of $1-$SignSGD and $\infty-$SignSGD when generalizing to the receiver $Rx_{(1-19)}$ with different noise variances. It can be seen that as the noise variance increases, the classification accuracy increases and then decreases, reaching a maximum close to the performance without gradient compression, both outperforming signSGD. Fig.~\ref{cost} plots the testing accuracy of all methods versus the accumulated number of bits transmitted from the clients to the server. One can see that the three gradient compression methods significantly enhance communication efficiency, with minimal performance loss for uncompressed FedRIEI.
	
	\begin{table*}[ht]
		\renewcommand\arraystretch{1.3}
		\centering
		\caption{Comparison of recognition accuracy (\%) for different gradient compression methods.\label{compress}}
		%	\begin{adjustwidth}{-\extralength}{0cm}
			%		\newcolumntype{C}{>{\centering\arraybackslash}X}
			\begin{tabular}{cccccc}
				\hline
				\textbf{Training data receivers}& \textbf{Test data receiver}& \textbf{FedRIEI}& \textbf{SignSGD}& \textbf{$1-$SignSGD $\sigma=0.01$}& \textbf{$\infty-$SignSGD $\sigma=0.01$}\\
				\hline
				$Rx_{(1-1)}$, $Rx_{(1-19)}$, $Rx_{(7-7)}$, $Rx_{(8-8)}$  & $Rx_{(14-7)}$   & $76.54 \pm 0.26$   &  $75.71 \pm 1.04$&  $76.90 \pm 0.97$&  $74.80 \pm 0.74$  \\
				%			\hline
				$Rx_{(1-1)}$, $Rx_{(1-19)}$, $Rx_{(7-7)}$,$Rx_{(14-7)}$  & $Rx_{(8-8)}$   & $69.83 \pm 0.88$   &  $66.99 \pm 0.80$&  $67.63 \pm 0.43$&  $67.11 \pm 0.31$  \\
				%			\hline
				$Rx_{(1-1)}$, $Rx_{(1-19)}$, $Rx_{(8-8)}$, $Rx_{(14-7)}$  & $Rx_{(7-7)}$   & $75.07 \pm 0.85$   &  $70.10 \pm 1.69$&  $73.90 \pm 1.45$&  $73.35 \pm 0.67$  \\
				%			\hline
				$Rx_{(1-1)}$, $Rx_{(7-7)}$, $Rx_{(8-8)}$, $Rx_{(14-7)}$  & $Rx_{(1-19)}$  & $75.32 \pm 1.26$   &  $66.49 \pm 0.60$&  $67.40 \pm 0.88$&  $68.72 \pm 0.95$  \\
				%			\hline
				$Rx_{(1-19)}$, $Rx_{(7-7)}$, $Rx_{(8-8)}$, $Rx_{(14-7)}$  & $Rx_{(1-1)}$   & $75.32 \pm 0.55$   &  $75.36 \pm 0.77$&  $75.26 \pm 0.68$&  $75.55 \pm 0.58$  \\
				\hline
			\end{tabular}
			%	\end{adjustwidth}
		% \noindent{\footnotesize{* Tables may have a footer.}}
	\end{table*}
	
	\subsubsection{Effect of Sampling Rate}
\begin{figure}[]
		\centering
		\centerline{\includegraphics[width=9cm]{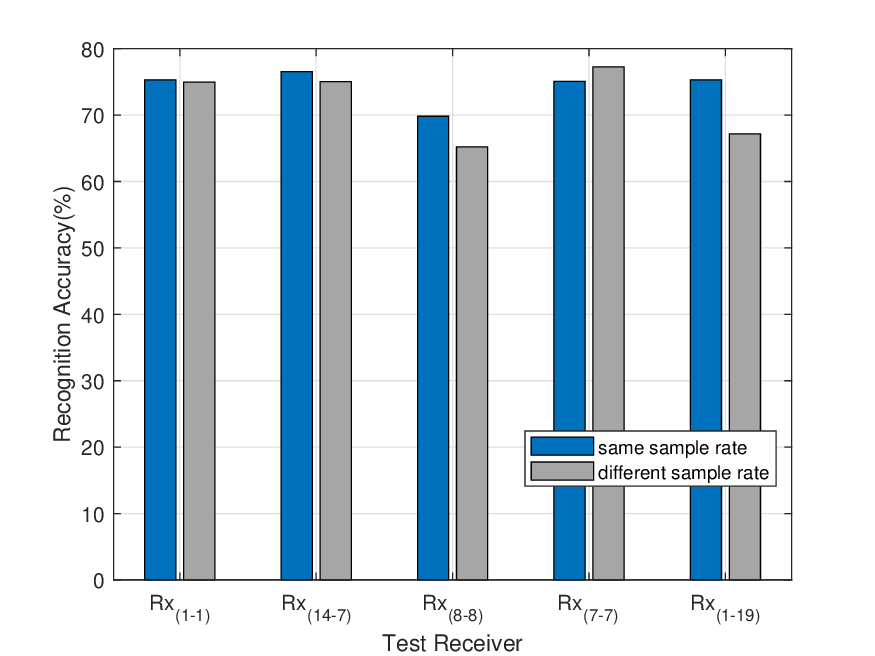}}
		\caption{The recognition accuracy of FedRIEI when four receivers are used for training and one for testing. ``Same sampling rate'' refers to all receivers operating at the same sampling rate, while ``different sampling rates'' indicates that each receiver operates at different sampling rate.}\label{sample}
   \end{figure}

	In practice, each receiver/client may record signals at different rates. To investigate the impact of different sampling rates on FedRIEI performance, we set $Rx_{(1-1)}$ as  the original sampling rate, while $Rx_{(1-19)}$ operated at 0.6 times, $Rx_{(7-7)}$ at 0.7 times, $Rx_{(8-8)}$ at 0.8 times, and $Rx_{(14-7)}$ at 0.9 times the original rate. To keep all the receivers having the same signal length, cubic spline interpolation is first performed at all the receivers except $Rx_{(1-1)}$. Fig.~\ref{sample} shows the recognition accuracy when four receivers are used for training and the remaining one for testing. For comparison,  the  recognition accuracy corresponding to  the same sampling rate at all the receivers is also plotted.
	It can be seen that varied sampling rates at the receivers have little impact on FedRIEI's recognition accuracy.

	%%%%%%%%%%%%%%%%%%%%%%%%%%%%%%%%%%%%%%%%%%
	\section{Conclusions and Discussions}\label{section:Conclusions}
	This paper has addressed the fingerprint contamination problem, which arises from different characteristics of receivers in Radio Frequency Fingerprint Identification (RFFI). Based on a theoretical analysis of the cross-receiver RFFI problem, we designed an RIEI model to achieve cross-receiver RFFI. The main idea of RIEI is to separate the receiver-related features from the emitter-related features by feature disentanglement. To this end, we have developed three specially designed losses, namely the cross-entropy loss, the mutual independence loss and the information entropy loss, to encourage the model to learn a receiver-independent feature space and thus address the challenge of generalizing to unseen receivers during the deployment phase. Moreover, considering practical limitation on centralized processing,  we also extend RIEI to a distributed scenario to eliminate the need for centralized servers to access local raw data. Simulation  results on real-world datasets demonstrated the effectiveness of RIEI and FedRIEI in suppressing the effect of receiver on RFFI.
	
	However, we should also note that for excessively high ISR (cf.~Fig.~\ref{interference}), the  recognition performance drops  significantly. Therefore, as a future work, it is worth investigating how  to further robustify the RIEI model under extremely interfered scenarios.

	\section{Acknowledgment}
The authors would like to  sincerely thank the anonymous reviewers for their helpful and insightful comments.

		\bibliographystyle{IEEEtran}
		\bibliography{ref}

	\vspace{-30pt}
	\begin{IEEEbiography}[{\resizebox{.94in}{!}{\includegraphics{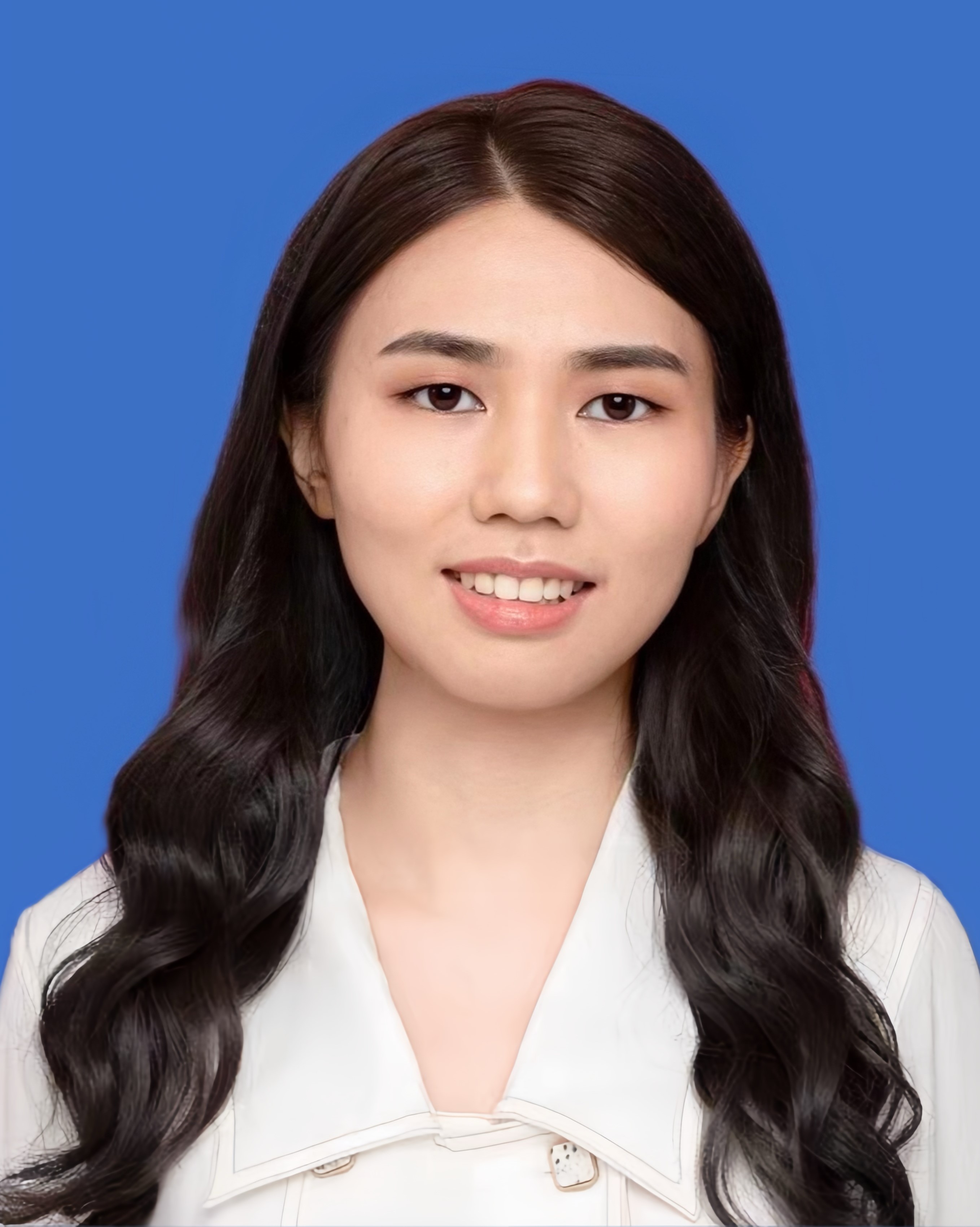}}}]
		{Ying Zhang} received the B.Eng. degree from Chongqing University of Posts and Telecommunications, Chongqing, China, in 2014, and the M.S. degree in electronic and communication engineering from the University of Electronic Science and Technology of China (UESTC), Chengdu, China, in 2017. She is currently pursuing the Ph.D. degree at the School of Information and Communication Engineering, UESTC. Her research interests include wireless communications and signal processing.
	\end{IEEEbiography}
%	\vspace{-30pt}
	\begin{IEEEbiography}[{\resizebox{.94in}{!}{\includegraphics{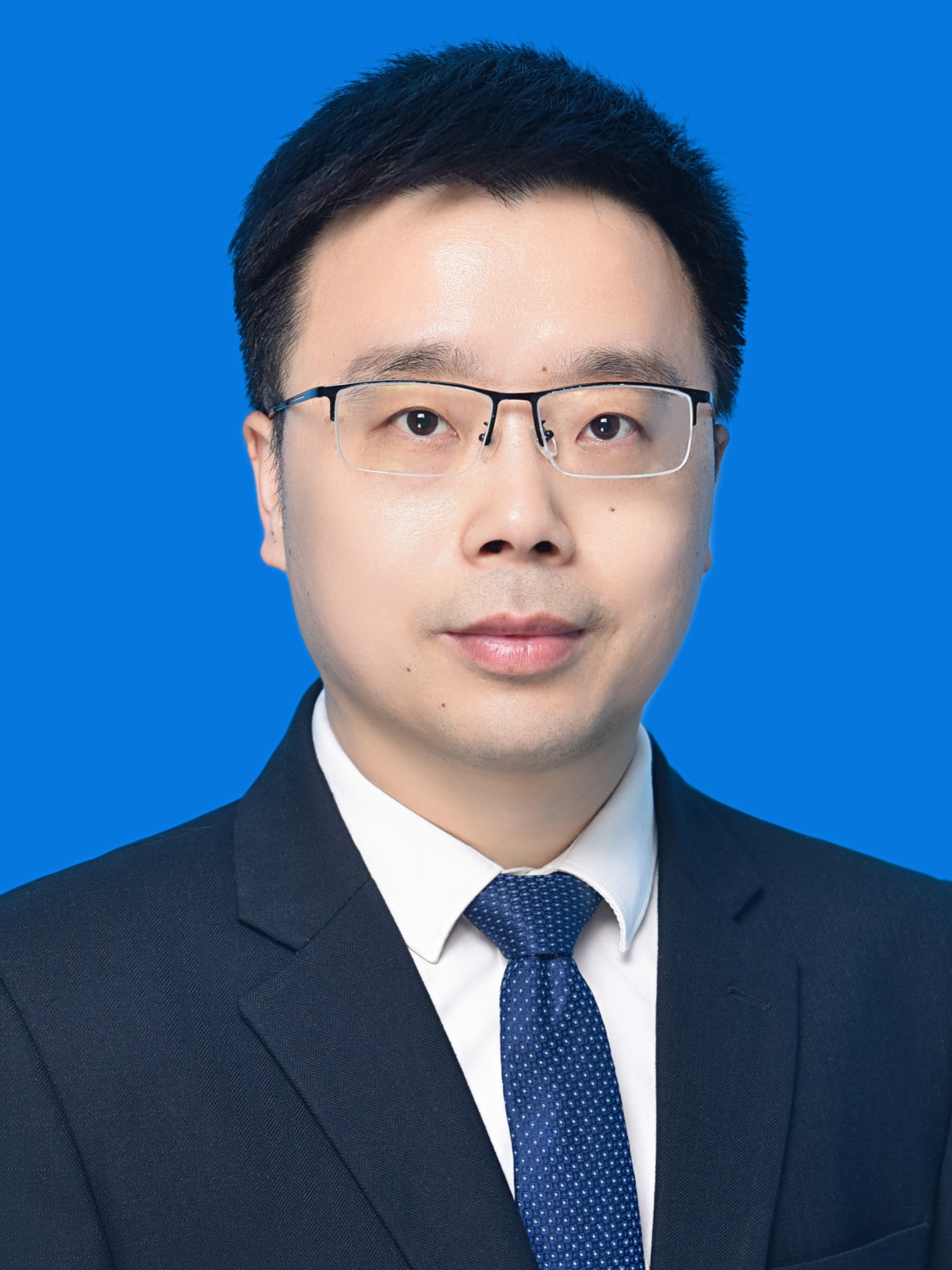}}}]
		{\bf Qiang Li} received the B.Eng. and M.Phil. degrees in communication and information engineering from the University of Electronic Science and Technology of China (UESTC), Chengdu, China, and the Ph.D. degree in electronic engineering from the Chinese University of Hong Kong (CUHK), Hong Kong, in 2005, 2008, and 2012, respectively. He was a Visiting Scholar with the University of Minnesota, and a Research Associate with the Department of Electronic Engineering and the Department of Systems Engineering and Engineering Management, CUHK. Since November 2013, he has been with the School of Information and Communication Engineering, UESTC, where he is currently a Professor. His recent research interests focus on machine learning and intelligent signal processing in wireless communications. He received the First Prize Paper Award in the IEEE Signal Processing Society Postgraduate Forum Hong Kong Chapter in 2010, a Best Paper Award of IEEE PIMRC in 2016, and the Best Paper Award of the IEEE SIGNAL PROCESSING LETTERS in 2016.
	\end{IEEEbiography}
	\vspace{-240pt}
	\begin{IEEEbiography}[{\resizebox{.94in}{!}{\includegraphics{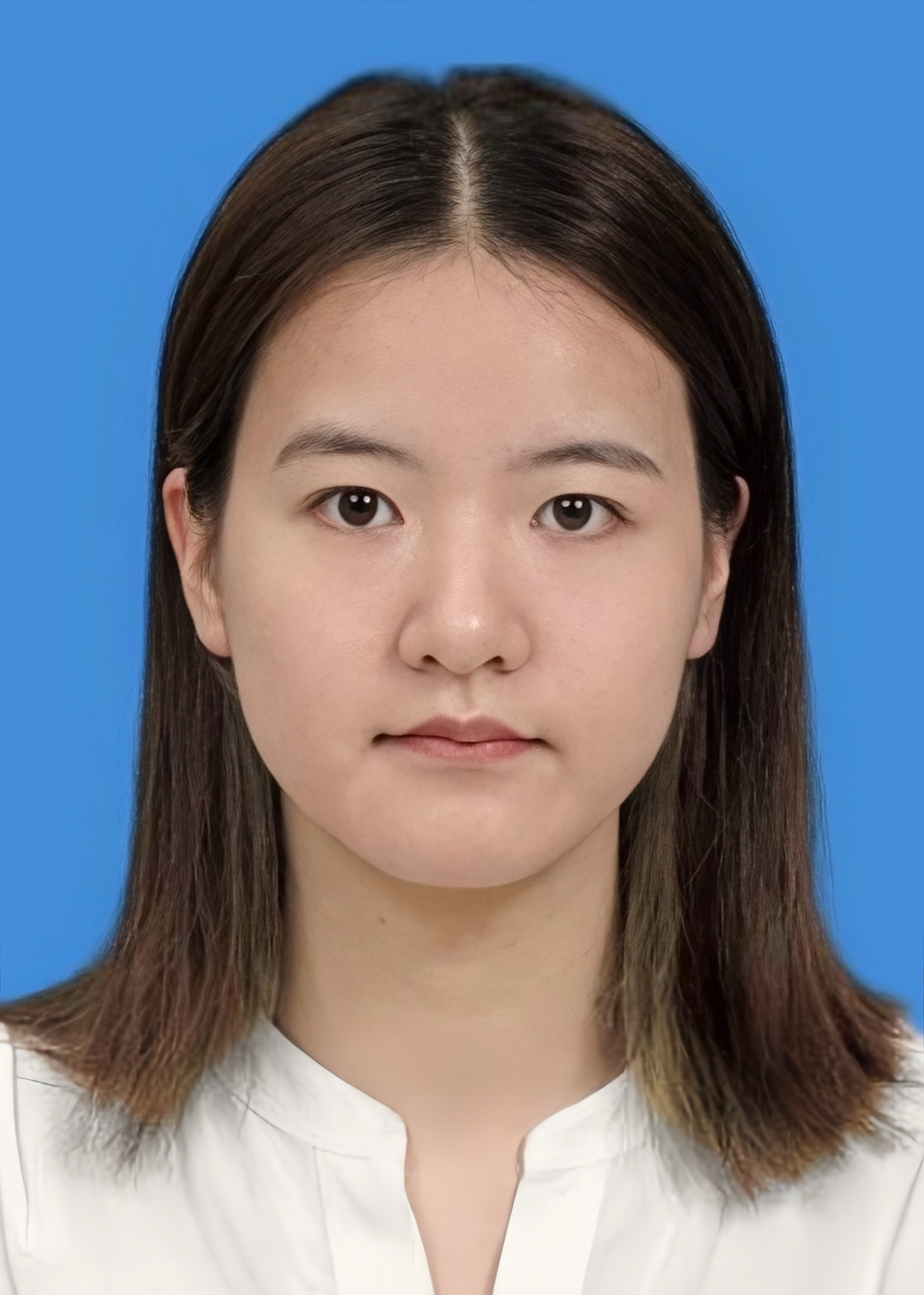}}}]
		{\bf Hongli Liu} received the B.Eng. degree from the Southwest University, Chongqing, China, in 2022. She is currently working toward the Ph.D. degree at the School of Information and Communication Engineering, University of Electronic Science and Technology of China, Chengdu, China. Her research interests are mainly on wireless communications, massive MIMO systems, and integrated sensing and communication.
	\end{IEEEbiography}
	\vspace{-240pt}
	\begin{IEEEbiography}[{\resizebox{.94in}{!}{\includegraphics{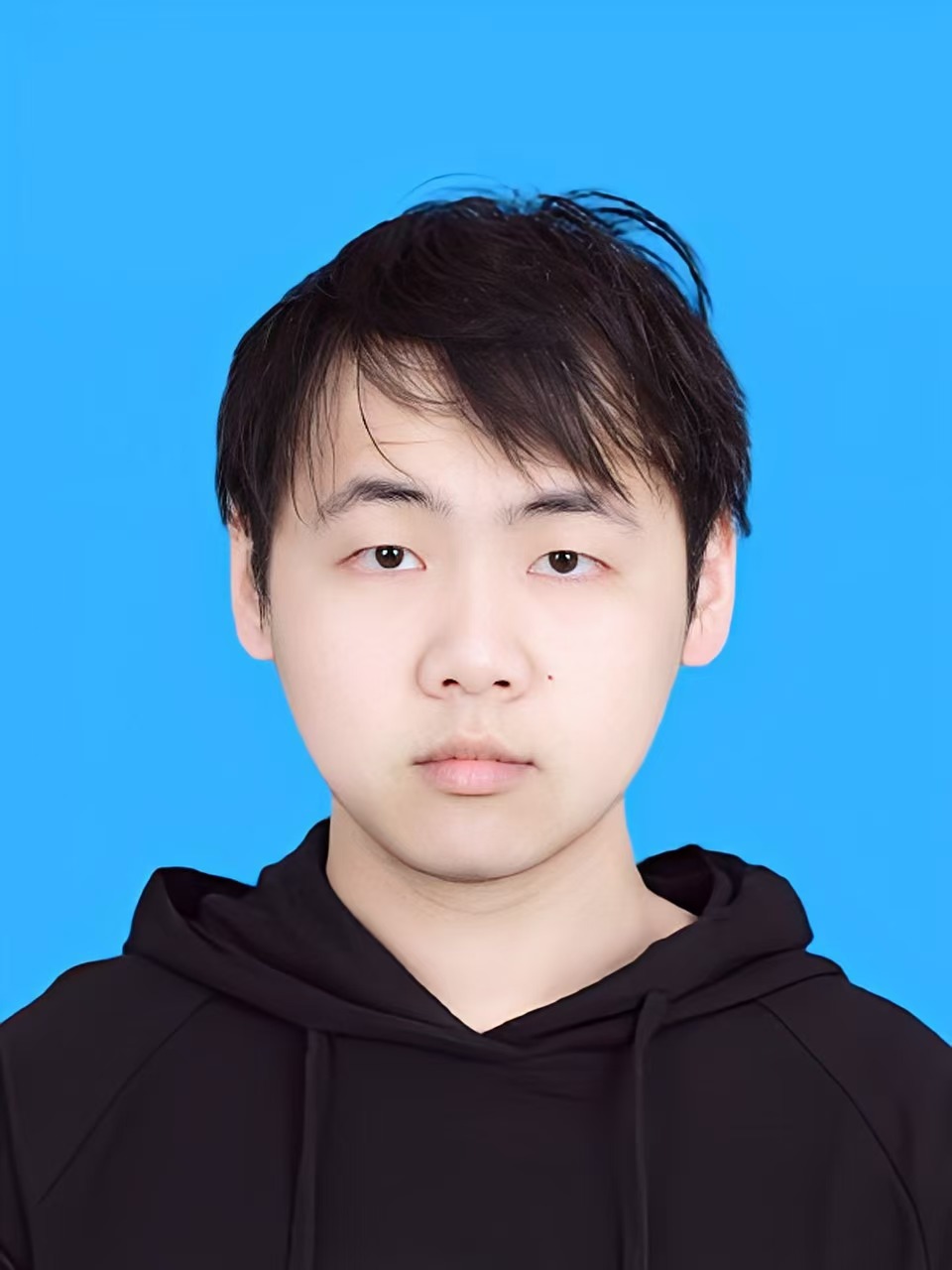}}}]
		{\bf Liu Yang} received the B.Eng. degree from the University of Electronic Science and Technology of China (UESTC), Chengdu, China, in 2020. He is currently pursuing the Ph.D. degree with the School of Information and Communication Engineering, UESTC. His current research interests include signal processing and radio frequency fingerprint identification.		
	\end{IEEEbiography}
		\vspace{-240pt}
	\begin{IEEEbiography}[{\resizebox{.94in}{!}{\includegraphics{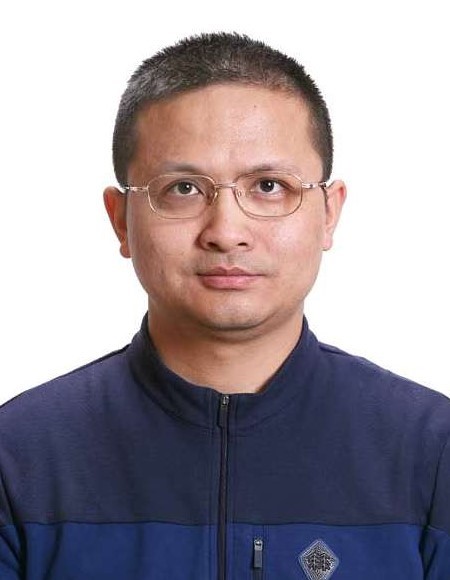}}}]
		{\bf Jian Yang} received the B.S. and M.S. degrees in communication and information systems from Information Engineering University, Zhengzhou,China, in 2003 and 2006, respectively. After his graduation, he joined the Northern Institute of Electronic Equipment of China, Beijing, China. He is currently pursuing the Ph.D. degree with the School of Cyberspace Science and Technology, Beijing Institute of Technology, Beijing, China. He is mainly interested in communication signal processing.		
	\end{IEEEbiography}

	% \vfill
	
\end{document}